\documentclass[tradiabstract,longauth]{aa}  
\usepackage[osf]{libertine} 
\usepackage[T1]{fontenc}
\usepackage{textcomp}
\usepackage[varqu,varl]{inconsolata} 
\usepackage{amsmath}
\usepackage[cal=boondoxo]{mathalfa} 
\usepackage[varg]{txfonts}
\usepackage{graphicx}
\usepackage{lscape}
\usepackage{url}
\usepackage[draft]{hyperref}

\usepackage{natbib}
\usepackage{multirow}
\usepackage{color}
\usepackage{amssymb}
\usepackage[dvipsnames]{xcolor}
\usepackage{enumerate}
\usepackage{placeins}
\usepackage{soul}
\usepackage{subfig}

\pdfoutput=1

\newcommand{\jp}{J-PAS}
\newcommand{\js}{J-spectra}
\newcommand{\mjp}{miniJPAS}

\def\sext{\texttt{SExtractor}}

\def\photozbest{\texttt{PHOTOZ}}

\def\magauto{\texttt{MAG\_AUTO}}

\def\magpsfcor{\texttt{MAG\_PSFCOR}}

\def\baysea{\texttt{BaySeAGal}} 
\def\muff{\texttt{MUFFIT}} 
\def\alstar{\texttt{AlStar}} 
\def\tgas{\texttt{TGASPEX}}

\def\rb{$r_\mathrm{SDSS}$}
\def\gb{$g_\mathrm{SDSS}$}
\def\ib{$i_\mathrm{SDSS}$}
\def\uja{$u_\mathrm{JAVA}$}
\def\ujp{$u_\mathrm{JPAS}$}

\newcommand\ewha{$EW(\mathrm{H}\alpha)$}
\newcommand\logMt{\log M_\star}
\newcommand\logM{$\logMt$}

\bibpunct{(}{)}{;}{a}{}{,}
\graphicspath{{Figures/}}

\begin{document}

\title{The \mjp \ survey:}
\subtitle{The galaxy populations in the most massive cluster in \mjp, mJPC2470-1771}

\authorrunning{ J.E. Rodríguez Martín and J-PAS collaboration}
\titlerunning{Cluster mJPC2470-1771}

\author{J.E. Rodríguez Martín\inst{\ref{IAA}}
\and 
R.~M.~Gonz\'alez Delgado\inst{\ref{IAA}}
\and 
G.~Mart\'inez-Solaeche\inst{\ref{IAA}}
\and
L.~A.~D\'iaz-Garc\'ia\inst{\ref{IAA}}
\and
A.~de Amorim\inst{\ref{UFSC}}
\and
R.~Garc\'ia-Benito\inst{\ref{IAA}}
\and
E.~P\'erez\inst{\ref{IAA}}
\and
R.~Cid Fernandes\inst{\ref{UFSC}}
\and
E.~R.~Carrasco\inst{\ref{GEMINI}}
\and
M.~Maturi\inst{\ref{ZFA}, \ref{ITP}}
\and
A.~Finoguenov\inst{\ref{DPUH}}
\and
P.~A.~A.~Lopes\inst{\ref{OV}}
\and
A.~Cortesi\inst{\ref{OV}}
\and
G.~Lucatelli\inst{\ref{IMEF}}
\and
J.~M. Diego\inst{\ref{IFC}}
\and
A.~L.~Chies-Santos\inst{\ref{UFRGS}, \ref{SAO}}
\and
R.~A.~Dupke\inst{\ref{ON},\ref{DAUM},\ref{DPA}}
\and
Y.~Jim\'enez-Teja\inst{\ref{IAA}}
\and
J.~M.~V\'ilchez\inst{\ref{IAA}}
\and
L.~R.~Abramo\inst{\ref{USP-IF}}
\and
J.~Alcaniz\inst{\ref{UFRGS}, \ref{ON}}
\and
N.~Ben\'itez\inst{\ref{IAA}}
\and
S.~Bonoli\inst{, \ref{DIPC}, \ref{CEFCA},\ref{BFS}}
\and
A.~J.~Cenarro\inst{\ref{CEFCA}, \ref{CEFCA2}}
\and
D.~Crist\'obal-Hornillos\inst{\ref{CEFCA}}
\and
A.~Ederoclite\inst{\ref{USP}}
\and
A.~Hern\'an-Caballero\inst{\ref{CEFCA}}
\and
C.~L\'opez-Sanjuan\inst{\ref{CEFCA}, \ref{CEFCA2}}
\and
A.~Mar\'in-Franch\inst{\ref{CEFCA}, \ref{CEFCA2}}
\and
C.~Mendes de Oliveira\inst{\ref{USP}}
\and
M.~Moles\inst{\ref{IAA},\ref{CEFCA}}
\and
L.~Sodr\'e Jr.\inst{\ref{USP}}
\and
K.~Taylor\inst{\ref{USP}}
\and
J.~Varela\inst{\ref{CEFCA},\ref{CEFCA2}}
\and
H.~V\'azquez Rami\'o\inst{\ref{CEFCA},\ref{CEFCA2}}
\and
I.~M\'arquez\inst{\ref{IAA}}
}

\institute{
Instituto de Astrof\'{\i}sica de Andaluc\'{\i}a (CSIC), P.O.~Box 3004, 18080 Granada, Spain\newline \email{rosa@iaa.es}\label{IAA}
\and 
Departamento de F\'{\i}sica, Universidade Federal de Santa Catarina, P.O.~Box 476, 88040-900, Florian\'opolis, SC, Brazil\label{UFSC}
\and
Gemini Observatory/NSF’s NOIRLab, Casilla 603, La Serena, Chile\label{GEMINI}
\and
Zentrum für Astronomie, Universität Heidelberg, Philosophenweg 12, D-69120 Heidelberg, Germany\label{ZFA}
\and
Institut für Theoretische Physik, Universität Heidelberg, Philosophenweg 16, D-69120 Heidelberg, Germany\label{ITP}
\and
Department of Physics, University of Helsinki, Gustaf Hällströmin katu 2, FI-00014 Helsinki, Finland\label{DPUH}
\and
Observat\'orio do Valongo, Universidade Federal do Rio de Janeiro, 20080-090, Rio de Janeiro, RJ, Brazil\label{OV}
\and
Instituto de Matematica Estatistica e Fisica, Universidade Federal do Rio Grande (IMEF–FURG), Rio Grande, RS, Brazil\label{IMEF}
\and
Instituto de F\'isica de Cantabria (CSIC-UC). Avda. Los Castros s/n. 39005, Santander, Spain\label{IFC}
\and
Departamento de Astronomia, Instituto de F\'isica, Universidade Federal do Rio Grande do Sul (UFRGS), Av.~Bento Gonçalves 9500, Porto Alegre, R.S, Brazil\label{UFRGS}
\and
Shanghai Astronomical Observatory, Chinese Academy of Sciences, 80 Nandan Rd., Shanghai 200030, China\label{SAO}
\and
Observat\'orio Nacional, Minist\'erio da Ciencia, Tecnologia, Inovaç\~ao e Comunicaç\~oes, Rua General Jos\'e Cristino, 77, S\~ao Crist\'ov\~ao, 20921-400, Rio de Janeiro, Brazil\label{ON}
\and
Department of Astronomy, University of Michigan, 311 West Hall, 1085 South University Ave., Ann Arbor, USA\label{DAUM}
\and
Department of Physics and Astronomy, University of Alabama, Box 870324, Tuscaloosa, AL, USA\label{DPA}
\and
Instituto de F\'isica, Universidade de S\~ao Paulo, Rua do Mat\~ao 1371, CEP 05508-090, S\~ao Paulo, Brazil\label{USP-IF}
\and
Donostia International Physics Center (DIPC), Manuel Lardizabal Ibilbidea 4, San Sebasti\'an, Spain\label{DIPC}
\and
Centro  de  Estudios  de  F\'isica  del  Cosmos  de  Arag\'on  (CEFCA),  Plaza San Juan 1, E-44001, Teruel, Spain\label{CEFCA}
\and
Centro  de  Estudios  de  F\'isica  del  Cosmos  de  Arag\'on  (CEFCA), Unidad Asociada al CSIC, Plaza San Juan 1, E-44001, Teruel, Spain\label{CEFCA2}
\and
Ikerbasque, Basque Foundation for Science, E-48013 Bilbao, Spain\label{BFS}
\and
Instituto de F\'isica, Universidade Federal da Bahia, 40210-340, Salvador, BA, Brazil\label{UFB}
\and
Universidade de S\~{a}o Paulo, Instituto de Astronomia, Geof\'isica e Ci\^encias Atmosf\'ericas, R. do Mat\~{a}o 1226, 05508-090, S\~{a}o Paulo, Brazil\label{USP}
}
\date{\today}

\abstract{The Javalambre-Physics of the Accelerating Universe Astrophysical Survey (J-PAS) is a photometric survey that will scan thousands of square degrees of the sky. It will use 54 narrow-band filters, combining the benefits of low-resolution spectra and photometry. The \mjp \ is a 1 deg$^2$ survey that uses \jp \ filter system with the Pathfinder camera. We study mJPC2470-1771, the most massive cluster detected in \mjp. We study the stellar population properties of the members, their star formation rates (SFR), star formation histories (SFH), the emission line galaxy (ELG) population, the spatial distribution of these properties, and the effect of the environment on them. This work shows the power of  \jp \ to study the role of environment in galaxy evolution.  
 We use a spectral energy distribution (SED) fitting code to derive the stellar population properties of the galaxy members: stellar mass, extinction, metallicity, $(u-r)_{\mathrm{res}}$ and $(u-r)_{\mathrm{int}}$ colours, mass-weighted age, the SFH, parametrised by a delayed-$\tau$ model ($\tau$, $t_0$), and SFRs. Artificial Neural Networks are used for the identification of the ELG population through the detection of H$\alpha$, [NII], H$\beta$, and [OIII] nebular emission. We use the Ew(H$\alpha$)-[NII] (WHAN) and [OIII]/H$\alpha$-[NII]/H$\alpha$ (BPT) diagrams to separate them into star-forming galaxies, and AGNs. 
We find that the fraction of red galaxies increases with the cluster-centric radius; and at $0.5$~R$_{200}$ the red and blue fractions are both equal. The redder, more metallic and massive galaxies tend to be inside the central part of the cluster, 
while blue, less metallic and less massive galaxies are mainly located outside of the inner $0.5$~R$_{200}$. We select 49 ELG, $65.3$~\% of the them are probably star forming galaxies, and they are dominated by blue galaxies. $24$~\% are likely to have an AGN (Seyfert or LINER galaxies). The rest are difficult to classify and are most likely composite galaxies. 
These galaxies are red, and their abundance decreases with the cluster-centric radius; in contrast, the fraction of star forming galaxies increases outwards up to $R_{200}$. 
Our results are compatible with an scenario where galaxy  members  were formed roughly at the same epoch, but blue galaxies  have had more recent star formation episodes, and they are quenching from inside-out of the cluster centre. The spatial distribution of red galaxies and their properties suggest that they were quenched prior to the cluster accretion or an earlier cluster accretion epoch. AGN feedback and/or mass might also be intervening in the quenching of these galaxies.
}




\maketitle

\section{Introduction}
\label{sec:Introduction}
Galaxies in clusters interact with each others and with the intracluster medium through processes such as ram-pressure stripping \citep{Gunn1972}, tidal stripping \citep{Malumuth1984} or harassment \citep{Moore1996}. These processes affect their star formation and evolution, and can lead to a greater presence of massive galaxies in dense environments and lower star formation rates (SFRs) in such regions \citep[e.g.][]{Lewis2002, Gomez2003, Baldry2006}. Therefore, galaxies in clusters are a great laboratory for studying the role of environment in galaxy evolution.

Certainly, interactions within galaxy clusters play a relevant role on the transformation of galaxies (\citealt{boselli2006}). Since the pioneering work by  \cite{Dressler1980} it is well-known that there is a morphology--density relation that may imply a connection between dense environments and the transformation and evolution of galaxies. This relation shows that as local galaxy density increases, so does the fraction of early-type galaxies, and the fraction of spirals decreases. \citet{Dressler1980} explains this relation as a reflection of the time scale of the formation of the disc of galaxies. This morphology--density relation has been confirmed by many works both at the nearby universe \citep[e.g.][]{Cappellari2011, fogarty2014} and at higher redshift \citep[e.g.][]{muzzin2012}. This relation could be the result of galaxy-galaxy merging processes and ram-pressure stripping, since they can produce the formation of a spheroidal component, leading to the morphological transformation of late to early type galaxies \citep{Boselli2008, Rijcke2010, Joshi2020, Peschken2020, Janz2021}.

The effect of dense environments can be also seen in the properties of galaxy populations.  Density strongly affects the stellar mass distribution and, at fixed stellar mass, the star formation and the nuclear activity depends on the density too, but the structural parameters were independent of the environment \citep{Kauffmann2004}. \cite{Balogh2004} showed that, at fixed luminosity, the mean $(u-r)$ colour of red and blue galaxies is almost independent of the environment, but the fraction of red galaxies increases with density. They propose that the transformations from blue to red must occur very rapidly (in this case the process is known as `quenching'), or at high redshift. In this line, \cite{Bower1990} results also point out that galaxies in denser environments are older on average, meaning that galaxies in denser environments have their star formation truncated at earlier epochs, as opposed to galaxies in less dense environments. This age--density relation is also seen in red sequence galaxies in the work by \cite{Cooper2010MNRAS}, showing also a weak correlation between metal-rich galaxies and denser environments. This age-density relation is supported by several other works \citep[e.g. ][]{Trager2000, Thomas2005,Clemens2006, Bernardi2006, Smith2008}

Clusters, in particular those formed at more recent times, are also dynamically in-mature structures, that have doubled their mass since $z\sim 0.5$ \citep{boylan2009, gao2012}.  The accretion times for $z = 0$ cluster members are quite extended, with $\sim 20$~\% of them incorporated into the cluster halo more than 7~Gyr ago and  $\sim 20$~\% within the last 2 Gyr  (\citealt{berrier2009}). Thus, the galaxy population in clusters have evolved rapidly since $z\sim$0.5, with the accretion of star forming galaxies into the cluster and their transformation into early-type red galaxies.

The so-called Butcher-Oemler effect \citep{ButcherOemler1978, ButcherOemler1984} also reflects the evolutionary nature of clusters. It shows that the fraction of blue galaxies is larger for clusters at higher redshift  (e.g. \citealt{balogh2000,Ellingson2001,diaferio2001}). Moreover, they find that blue galaxies are mostly located outside the cluster cores and that the effect is not significant for distances larger than $0.5$~R$_{200}$. In fact, passive galaxies are mainly located in the virialised regions while the emission line galaxies are more common in the outskirts of the clusters \citep{Haines2012, Haines2015, Noble2013, Noble2016, Mercurio2021}. In contrast,  the Faint Infrared Grism Survey \citep[FIGS, ][]{FIGS} find that [OIII] emitters are more common close to groups, but there is no evidence of a relation between SFR and local galaxy density \citep{Pharo2020}.

Quenching is an important effect related not only to the environment, but also to the galaxy mass. \cite{Peng2010} separate the effects of the mass and the environment in stopping the star formation and consider quenching as a combination of both effects (mass-quenching and environment-quenching). Other works have also shown that certain processes that are in some way related to the galaxy stellar mass can suppress the star formation in galaxies independently of the environment \citep[see e.g.][]{ Peng2012, ArcilaOsejo2019, Contini2020, Guo2021}. In fact, AGN feedback can play a relevant role by heating the infalling gas preventing for further star formation in the galaxy  \citep{dimatteo2005, Fabian2012, Mcnamara2012}, although it remains openly discussed \citep{Esposito2022, Wang2022}. The central velocity dispersion is  correlated with the mass of the central black hole of galaxies, so it is  connected the AGN feedback, and it has been shown to play a crucial role in quenching \citep[see e.g.][]{Bluck2020, Brownson2022}.

Some environmental processes can temporarily enhance the star formation due to the inflow of gas toward the central part (e.g. galaxy-galaxy interactions) and/or the compression of the gas \citep[e.g. ram pressure stripping][]{Joseph1985, Park2009, Ellison2013, RuizLara2020, Boselli2021, Mazzi2021, Lizee2021}. Nevertheless, these environment mechanisms eventually shut down the star formation by heating or removing the gas from galaxies  \citep{Altalo2015, Davies2015,Lisenfeld2017,Joshi2019}.

Alternatively, the halo mass is proposed as the main property that is causally linked to the rapid shut down of the star formation  \citep[see e.g.][]{Bluck2014, Woo2015, MonteroDorta2021} because the fraction of quenched galaxies is more correlated with the group/cluster halo mass  at a fixed M$_\star$ than with M$_\star$ at a fixed halo mass \citep{woo2013}. There exists a bimodality in the specific star formation rates (sSFR, the ratio of the SFR to the stellar mass) of satellite galaxies (those who fall into denser haloes) and they are more likely to be quenched than field galaxies \citep{Wetzel2012}. In addition, mass and environment quenching could be connected to massive halos that heat the cold-accreted gas preventing further star formation \citep{dekel2006}.  Furthermore, the quenching of satellites correlates not only with halo-mass, but also anticorrelates with the group/cluster radial distance  \citep{woo2013, Woo2017}. Nonetheless, IllustrisTNG simulations show that the dependence of the quenched galaxy fraction with the cluster-centric radius is also a function of the mass of halos. Thus,  although the fraction of quenched low-mass satellites in less massive halos  is higher closer to the centre, it is independent of the distance in massive halos \citep{Donnari2021}. However, after several decades of debate, the relevance of the different processes that can produce the quench of star formation and the transformation of galaxies in clusters is not clear yet. 

Photometric surveys with broad band filters, such as the Local Cluster Substructure Survey \citep[LoCuSS, ][]{Haines2015}, the Advance Large Homogeneous Area Medium Band Redshift Astronomical \citep[ALHAMBRA][]{Ascaso2015}, the VIMOS Public Extragalactic Redshift Survey \citep[VIPERS][]{Haines2017},  the Subaru Strategic Program with the Hyper Suprime-Cam \citep[HSC-SSP, ][]{Lin2017, Jian2018},  the Gemini Observations of Galaxies in Rich Environments \citep[GOGREEN][]{McNab2021}, have been essential for the detection and study of galaxy clusters. Surveys with narrow band filters at specific wavelengths to select emission lines \citep{Lin2017, Koyama2018, Hayashi2020}, and to identify infalling galaxies in the outskirt of clusters \citep{kodama2004}, have been very useful to identify the galaxy populations in clusters. However, these surveys could suffer from several problems related to the contamination of: 1) field interlopers due to the lack of precise redshifts for the galaxy cluster membership; 2) dusty star-forming galaxies in the cluster red galaxy population due to the non-correction of colours by extinction; 3) AGNs in the star-forming galaxy population. Further, the small FoV of some instruments, \citep[e.g. with Tunable Filters, as in][]{SanchezPortal2015, RodriguezdelPino2017} hinders the observation of the whole cluster or beyond the cluster centre. 

The Javalambre-Physics of the Accelerating Universe Astrophysical Survey (J-PAS; \citealt{Benitez2009,Benitez2014}) comes to overcome the problems associated with broad and narrow band photometric surveys. \jp{} will be very powerful in detecting galaxy clusters, and providing new clues for the understanding of the role of dense environment in quenching star formation in galaxies.  
\jp{} is a photometric survey that will  scan thousands of square degrees of the sky. With its 54 narrow-band filters (FWHM$\sim 145$~\AA, with a difference of $\sim 100~$\AA\ between the central wavelength of each one), plus two medium and four broadband filters, it will provide data comparable to very low resolution spectroscopy ($R\sim 60, \Delta \lambda \sim 100$~\AA). 

\jp{} is ideal for studies focused on the role of environment in galaxy evolution because it is very powerful in detecting galaxy clusters and groups \citep[see][]{Rosa2022}. It will be able to provide robust cluster or group detection based on accurate photometric redshifts \citep{HC2021}.The sensitivity of the survey allows us to easily observe the whole galaxy cluster memberships brighter than 22.5 in $r$ band, and  to study the quenching as a function of cluster-centric radius. \jp \ is ideal for SED fitting and to identify and characterise the blue and red galaxy populations \citep{Rosa2021}. Given its spectral coverage and resolution, it is able of identifying emission line objects, and to measure the lines H$\alpha$, [NII]$\lambda$6584, H$\beta$, and [OIII]$\lambda$5007 in clusters at $z<0.35$ \citep{Gines2021, Gines2022}. These lines are relevant to discriminate between the AGN and star-forming (SF) populations and to study their spatial distribution within the cluster. 

At present, there is available data using the \jp{} photometric system: miniJPAS \citep{Bonoli2020}. In this paper, we identify the galaxy populations to study the variation of galaxy properties as a function of the cluster-centric radius in the largest cluster detected in \mjp{}, mJPC2470-1771. The ultimate goal is to demonstrate the capability and the power of \jp \ for investigating the characterisation of galaxy populations in galaxy clusters, as well as the role of environment in quenching the star formation. This will allow us to shed light on the processes responsible for transforming blue and star-forming galaxies into red galaxies in this dense environments.

This paper is structured as follows.  In Sect.~\ref{sec:Data} we briefly summarise the  \mjp \ observations and calibrations and the selection of the cluster members; in Sect.~\ref{sec:SP} we describe the methods used to identify and study the stellar population properties of the galaxies and compare the results obtained with different photometries; in Sect.~\ref{sec:charact} we study the stellar population properties of these galaxy populations, we divide our ELG into star forming (SF) galaxies and galaxies with an active galactic nuclei (AGN) and we study the SFR of the cluster galaxies; in Sect.~\ref{sec:discussion} we discuss our results in terms of their spatial and radial distributions and in Sect.~\ref{sec:conclus} we summarise our results and present our conclusions.

Throughout this paper, we assume a Lambda cold dark matter ($\Lambda$CDM) cosmology with $h = 0.674$, $\Omega _{\mathrm{M}} = 0.315$, $\Omega _\Lambda = 0.685$, based on the latest results by the \cite{Planck2020}. This is the same cosmology used by \citet{Bonoli2020}. We use the AB magnitude system \cite{Oke1983}. We use the standard notation $M_{\Delta}$ for the mass enclosed within a sphere of radius $R_{\Delta}$, within which the mean overdensity equals $\Delta \times \rho_c(z)$ at a particular redshift $z$. That is, $M_{\Delta} = (4\pi \Delta/3)\rho_c(z)R_{\Delta}^{3}$.

\section{Data: \mjp \ }
\label{sec:Data}

\begin{figure}
\centering
\includegraphics[width=0.5\textwidth]{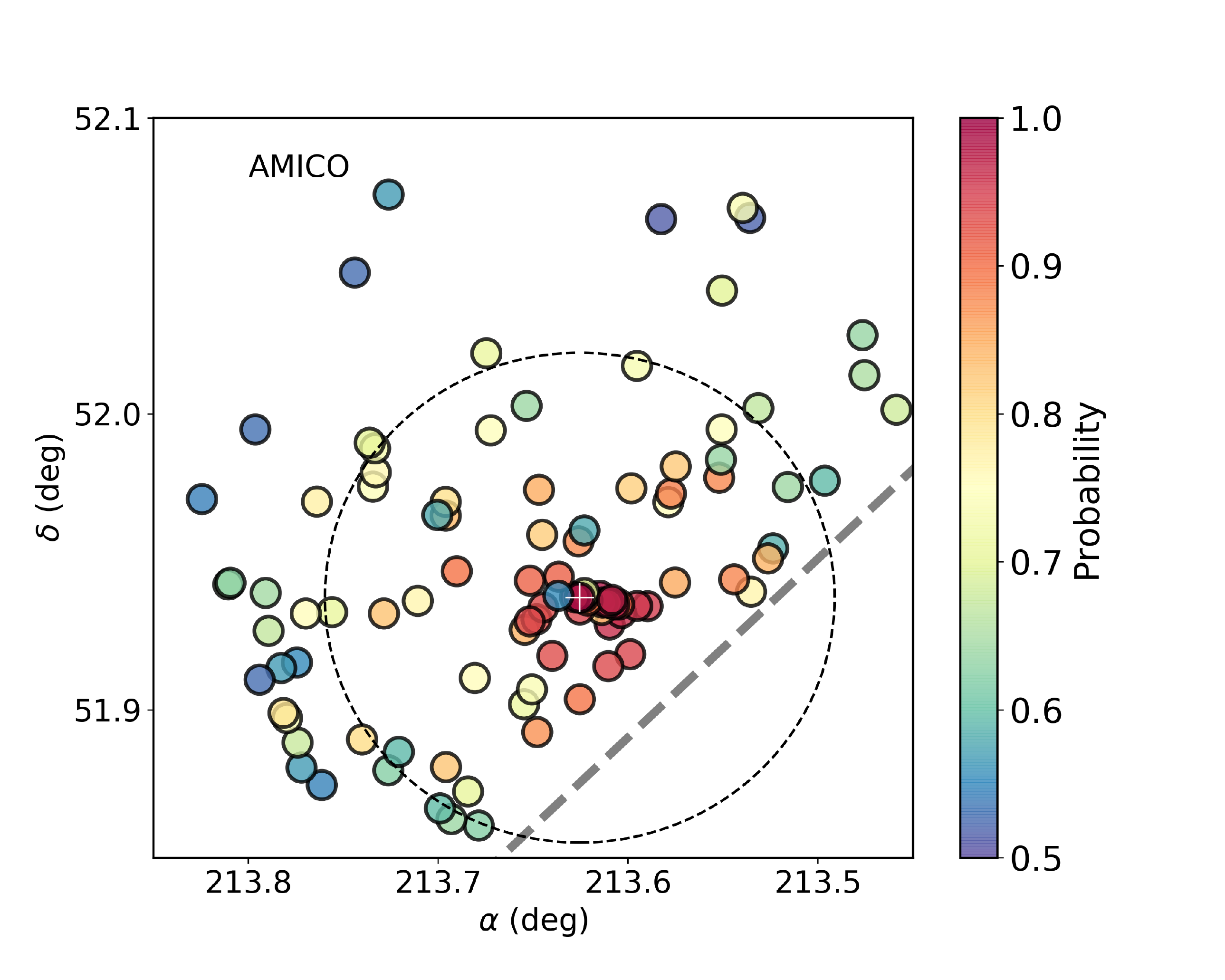}
\caption{Galaxy members of mJPC2470-1771 in the sky plane. The colour bar indicates the AMICO probability to be a member galaxy. The grey dashed line indicates the edge of the field of view of \mjp. The black dashed circle indicates the value of $R_{200}$. The white cross represents the position of the BCG.
}
\label{fig:mapProb}
\end{figure}

\begin{figure*}
\centering
\includegraphics[width=\textwidth]{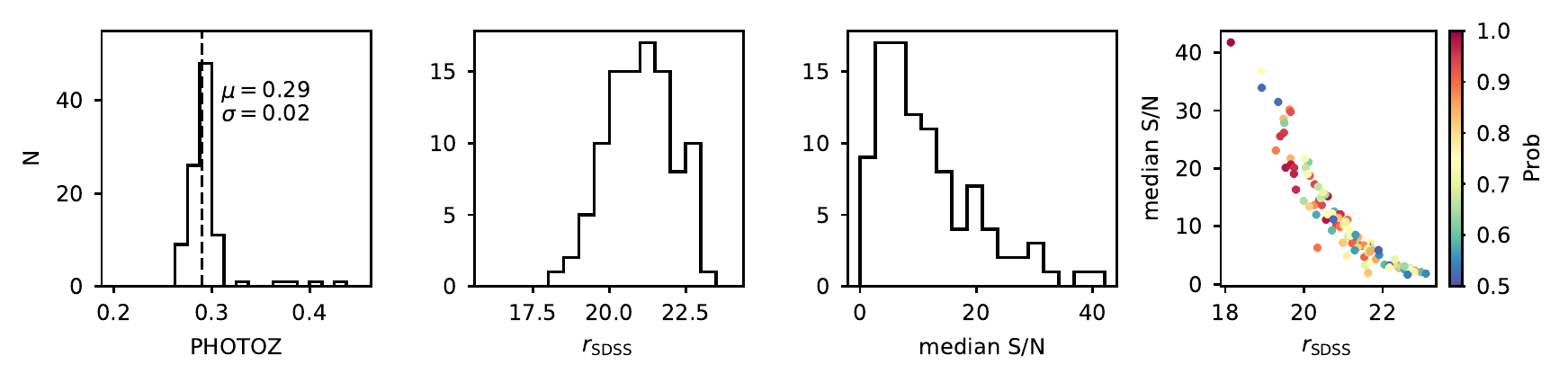}
\caption{First panel: Redshift (\photozbest{} ) distribution. The median ($\mu$) and standard deviation ($\sigma$) of \photozbest{} are written in the figure. The black dashed line represents the median of \photozbest{}. Second and third panel: Distributions of the \rb{} magnitude, and median S/N of the narrow-band filters. Fourth panel: median \magpsfcor \ S/N in the narrow-band filters as a function of \rb{} magnitude. Colour code indicates the probabilistic association given by AMICO.
}
\label{fig:histobs}
\end{figure*}

\subsection{Observations and calibration}
The \mjp \ survey \citep{Bonoli2020} is a $1~\mathrm{deg}^2$ imaging survey performed at the  Observatorio Astrofísico de Javalambre, \citep[OAJ, ][]{OAJ} using the 2.5m Javalambre Survey Telescope (JST/T250, \citealt{T250}), which provides a good image quality along the optical spectral range (3300--11000 \AA). The instrument used for the data acquisition is the  JPAS-Pathfinder camera. It has a single charge-coupled device (CCD)  with 9.2k$\times$9.2k pixel. The resulting field of view (FoV) is $0.27 ~\mathrm{deg}^2$ and the pixel scale is $0.23"~\mathrm{pixel}^{-1}$. The survey consists of four pointings along the AEGIS stripe \citep{AEGIS2007}. 

One of the J-PAS greatest strengths resides in its photometric system. It consists in 54 narrow-band (NB) filters with a full width at half maximum (FWHM)   of 145~\AA \ spaced by 100~\AA, covering the spectral range from 3780~\AA \ to 9100~\AA. There are two broader filters complementing these NB ones: \uja{}, a medium band filter with FWHM of 495~\AA \ and centred at 3497~\AA \ and J1007, a high-pass filter centred at 9316~\AA.  This system provides low resolution spectra ($R\sim60$) referred as \js \ and allows us to detect, identify and characterise the stellar population properties of galaxies up to $z \sim 1$ \citep{Rosa2021}. The filter system was originally optimised to accurately measure photometric redshifts (photo-$z$) for cosmological studies \citep{Benitez2009, Benitez2014, Bonoli2020}. In addition, four SSDS-like broadband filters are included: \ujp{}, \gb{}, \rb{}, and \ib{}. \rb{} is used as the reference detection band for the miniJPAS `dual-mode' catalogues. More information about the filter system can be found in \cite{Martin-Franch2012} and \cite{Bonoli2020}.

The area observed in \mjp \ overlaps with the AEGIS field, which is located in the North galactic hemisphere with coordinates ($\alpha$, $\delta$) = (215.00\degr, +53.00\degr). It is composed of four pointings covering a total area of 1~deg$^2$. The depth is deeper than 22~mag for filters with $\lambda<7500$~\AA \ and is $\sim 22$~mag for longer wavelengths. The data was processed by the Data Processing and Archiving Unit (UPAD, \citealt{Cristobal-Hornillos2014}) at Centro de Estudios de F\'isica del Cosmos de Arag\'on (CEFCA). Further details on the different processes involved (the processing of single images, the final coadded images, the PSF treatment, the photometry and its calibration and the masks) can be found in \cite{Bonoli2020}. Nonetheless, the data used in this work  was obtained with SExtractor dual-mode \citep{Bertin1996}. The photometric calibration is an adaptation of the methodology presented in \cite{Lopez-sanJuan2019}. All the images and catalogues are available through the CEFCA Web portal\footnote{\url{https://archive.cefca.es/catalogues}}, which also offers advanced tools for data searches, visualisations, and data queries \citep[see ][ and a future paper will be published, Civera et al., in prep.]{Civera2020}

\subsection{Identification of galaxy members}
\label{sec:galaxymembers}

\begin{figure*}
\centering
\includegraphics[width=0.8\textwidth]{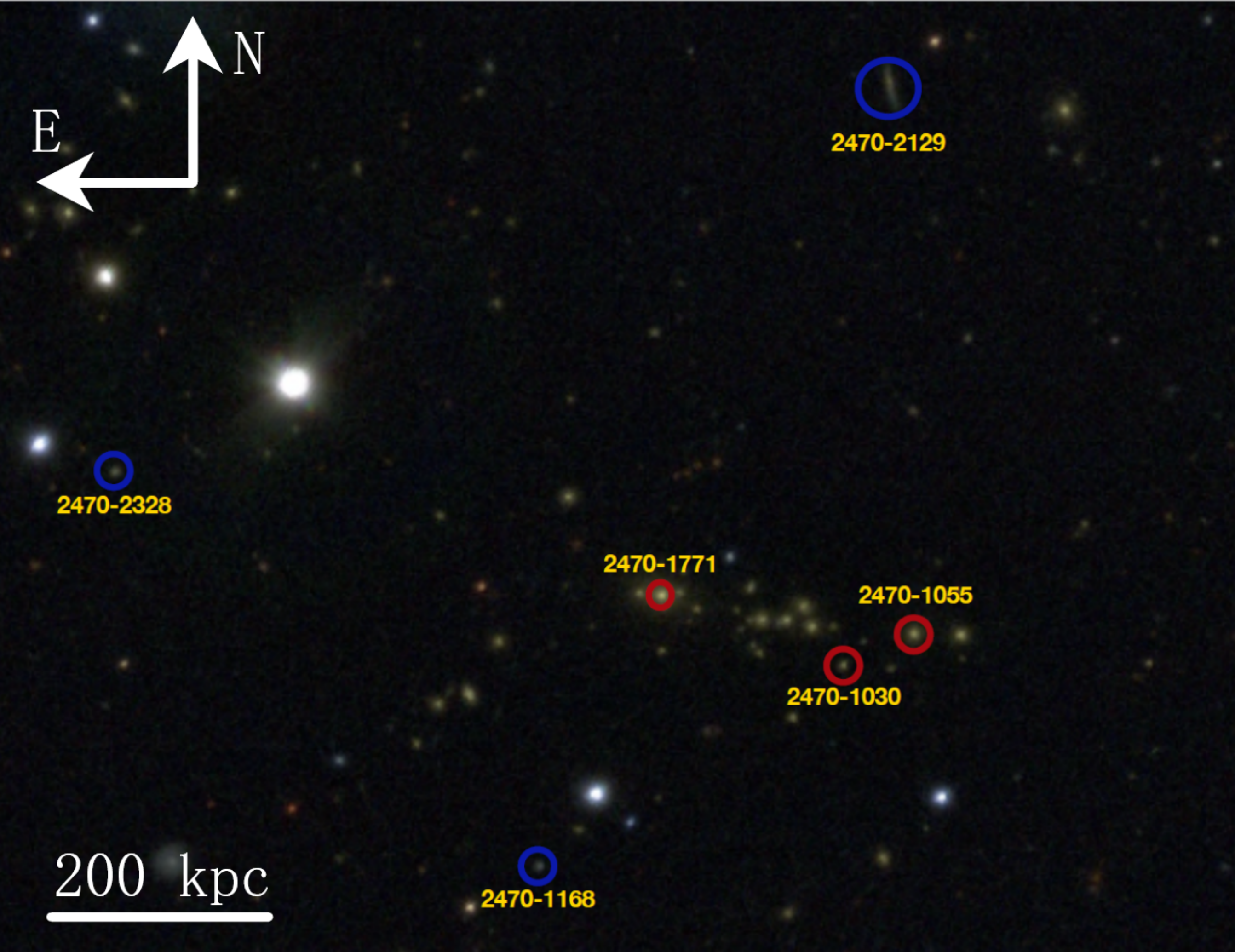}
\includegraphics[width=\textwidth]{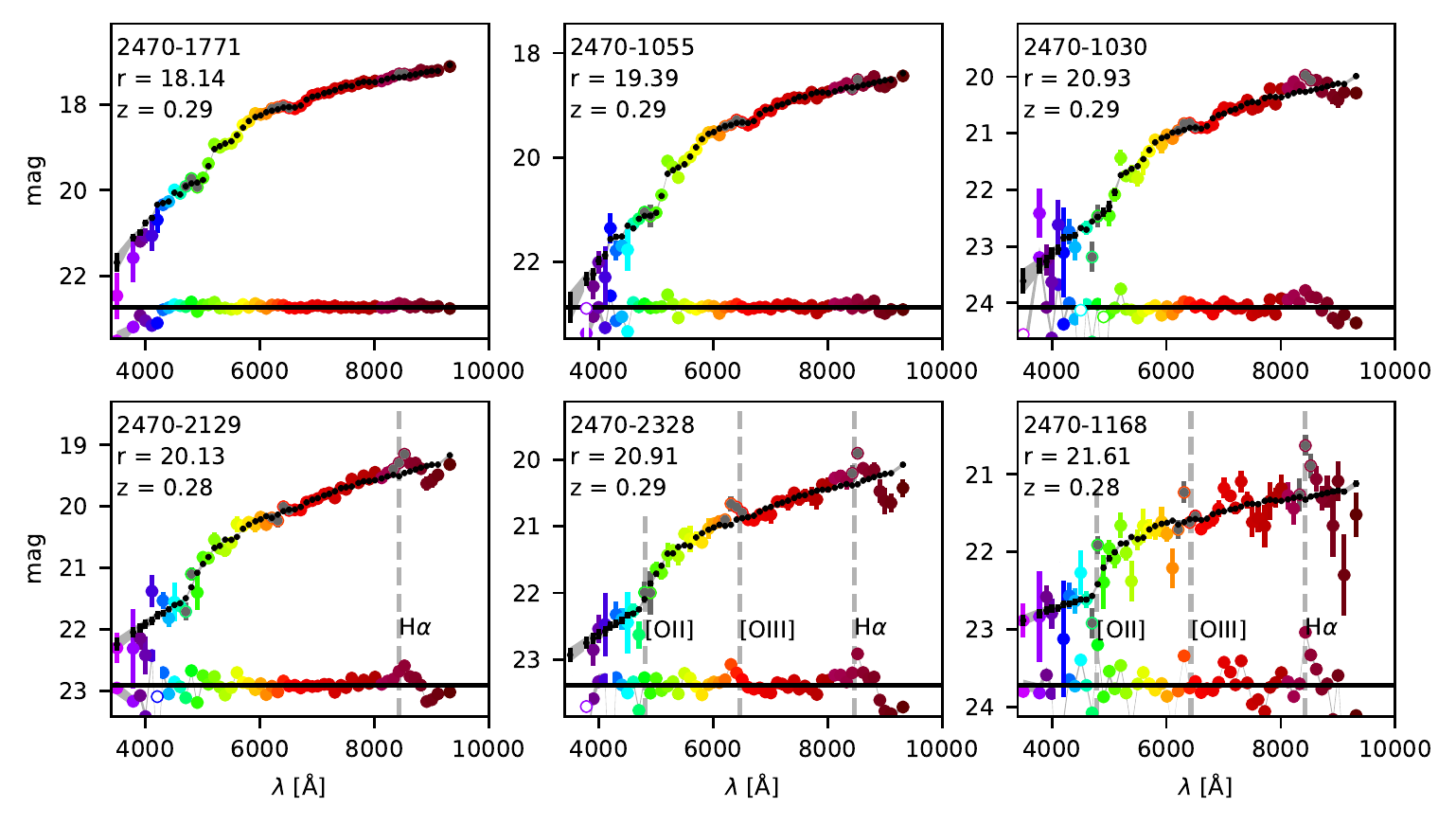}
\caption{Top panel: \mjp \ view of mJPC2470-1771. Prominent red and blue galaxies in the cluster are marked with red and blue circles respectively. The BCG corresponds to 2470-1771, with an spectroscopic redshift of $z=0.289$. Bottom panel: \magpsfcor \ \js \ of three red galaxies (top row) and three blue galaxies (bottom row) that are marked with circles in the top panel. 
The mean model fitted by \baysea \ is plotted as black points, and the grey band shows the magnitudes of the mean model $\pm$ one $\sigma$ uncertainty level.
The difference between the model and the best model fitted magnitudes are plotted as a small coloured points around the black bottom line. Masked filter (white coloured circles) and filters overlapping with the emission lines H$\alpha$, [NII], [OIII], H$\beta$ and [OII] (darker grey coloured circles) are not used in the fit. Vertical, grey dashed lines show the wavelengths corresponding to detectable emission lines.
}
\label{fig:jspectra}
\end{figure*}

The reference code for the detection of galaxy clusters in this work is Adaptive  Matched  Identifier of  Clustered  Objects (AMICO,  \citealt{2005A&A...436...37M, AMICO}). AMICO is an algorithm based on the Optimal Filtering technique \citep[see e.g.][]{Postman1996, Bellagamba2011, Bellagamba2018}. It uses a statistical description of the background noise and a template to characterise the signal of the clusters. The signal is defined as the product of the template and an amplitude, plus the noise component. It uses several inputs, mainly the galaxies' sky positions, their magnitudes and their redshifts to compute this amplitude and other parameters. Our parameter of interest is the association probability assigned to each galaxy, which represents the probability of the galaxy to be a member of the cluster. All the details and AMICO inputs used for making this catalogue of galaxy clusters  in miniJPAS will be detailed in a forthcoming work (Maturi et al., in prep.). 

For this work, we use the results from AMICO when using the \photozbest{} (the redshift corresponding to the maximum of the redshift probability density function, $z\mathrm{PDF}$ see \citealt{HC2021}). The choice of the redshift affects the galaxy members identified for the cluster, but for our purpose (the identification and characterisation of the galaxy populations in  mJPC2470-1771) there are no significant differences (see Appendix \ref{appendix:AMICOversions}). Therefore, we will use this catalogue, since it uses the same redshift as the one that was used for the SED-fitting analysis carried out by \cite{Rosa2021} using \baysea \ (see Sec. \ref{sec:method}).

The cluster, mJPC2470--1771, was identified as the most massive in \mjp \ by \cite{Bonoli2020}. The redshift of the cluster is $z=0.29$. In total there are 99 objects (see Fig.~\ref{fig:mapProb}) with  AMICO association probability higher than $0.5$ and brighter than 22.5 in the $r$ band. We identify the brightest cluster galaxy (BCG) as the galaxy with the highest luminosity in the \rb{} and the highest stellar mass. Its ID in the \mjp \ catalogue is 2470-1771 and its coordinates are $(\alpha, \delta)=(213.6254$\degr, $+51.9379$\degr). 

The catalogue of this cluster has been tested with the follow-up on Gemini/GMOS observation of 38 galaxies with probabilistic association larger than $0.5$ (Carrasco et al., in prep.). AMICO failed to classify as members of the cluster only for two galaxies, that have probabilities $0.58$ and $0.62$ in  the catalogue. Therefore, we estimate that AMICO classification only fails in 5\%  of the cases.

Using GMOS spectroscopy, we estimate that $R_{200}$ \footnote{$R_{200}$ is the radius where the mean mass overdensity is 200 times the critical density at the cluster's redshift.} is 1304 kpc, and the halo mass is  $M_{200}=3.3\times 10^{14} ~ M_\odot$. These estimates are based on the measurement of the velocity dispersion. The measurement took all the observed members and applied the Clean routine from \cite{CLEAN} on it. This routine iteratively estimates the velocity dispersion and removes the outliers based on the caustic profile. Velocity dispersion is estimated using MAD \citep{MAD}. 
Using this value of $R_{200}$, we see that some members of the cluster may be outside of our observing FoV (see  Fig. \ref{fig:mapProb}).

In fact, assuming that galaxies are symmetrically distributed up to $R_{200}$, we estimate that 9 galaxies that are between 0.5~$R_{200}$ and $R_{200}$ may not be included in our observing FoV. These galaxies represent only 12$\%$ of the sample; thus, any conclusion inside $R_{200}$ is robust. Outside $R_{200}$, the incompleteness could be higher, but it is difficult to evaluate, and could be up to $\sim$20-30$\%$ if the galaxy members show a circular symmetry. Thus, conclusions outside $R_{200}$ must be taken with caution. In any case, extensive properties, such as the stellar mass surface density, are corrected for this incompleteness.

The cluster has also been detected with other IDs, such as MaxBCG J213.62543+51.93786 \citep{Koester2007}, who find a detected richness of 19 (scaled richness 17); WHL J141430.1+515616 \citep{Wen2012}, who find a $R_{200}$ of $1.2$~Mpc, 30 objects inside $R_{200}$, and a richness of 34, or RM J141430.1+515616.5 \citep{Rozo2015a, Rozo2015b}. None of these works are dedicated to a specific study of the properties of the galaxy members of this cluster; moreover, they are very incomplete in their detection memberships. Thus, 
our work is the first and almost complete ($\sim$ 10$\%$ outside of the FOV) study inside $R_{200}$ for cluster memberships brighter than 22.5 (AB) in the $r$-band.


\subsection{Observational properties of galaxy members}
\label{sec:observationalproperties}
We have two different available photometries, \magauto \ and \magpsfcor. \magauto \ is provided by  \sext{} and estimates the total flux of the galaxy using an adaptive scaled aperture \citep[see ][and \sext{ manual for further details.}]{Bertin1996}. Using the same approach as \cite{Molino2019}, \magpsfcor \ aims to correct for the differences in PSF among different bands. It uses an aperture with the same shape as the Kron radius (smaller than the one used by \magauto) to provide robust colours determination \citep[see ][for further details]{Bonoli2020}. Due to their different apertures and extraction procedures \citep{Bonoli2020}, results between both may vary from one galaxy to another. In particular, \magauto \ uses a larger aperture, so it may include outer regions of the galaxy in the integration process, which tend to contain younger, blue stars. In fact, \cite{Rosa2021} compared the values of the stellar population properties obtained fitting the data from \magauto \ and \magpsfcor, finding that the main difference is that, on average, masses are $0.2$~dex larger in \magauto \ and rest frame colours are also bluer by $-0.09$~mag. Therefore, for our analysis, we will use \magpsfcor, given its better S/N. Its smaller aperture will also allow us to better detect the emission lines in the centre of the galaxies.

We first look at the observational properties of the galaxies (see Fig. \ref{fig:histobs}). The median measured redshift of the cluster's galaxies is $z=0.29$, with a standard deviation of $\sigma = 0.02$. There are four galaxies with redshift greater than 0.35. We looked at the $z\mathrm{PDF}$ of the galaxies in the galaxies in the cluster. These four galaxies are the only ones that show a multimodal distribution with peaks of similar amplitude (more than $\sim 50$~\% of the amplitude of the maximum peak). Due to their $z\mathrm{PDF}$ and their \photozbest{}, we decide to remove these four galaxies from our analysis.

The distribution of \rb \ peaks at around $\sim 21$~mag and most galaxies are brighter than $22.5$~mag. The number of galaxies in each bin steeply decreases with increasing median S/N, but the peak of the distribution is close to 10.  We see that brighter galaxies have a better S/N. Although there are galaxies with different probability and brightness, most of the galaxies with probability higher than $\sim 0.8$ have a magnitude brighter than 20 and a S/N higher than $\sim 11$. On the other hand, galaxies with probability lower than $\sim 0.6$ have a S/N higher than $\sim 10$. 

\section{Identification of the galaxy populations} 
\label{sec:SP}
The purpose of this section is to identify the red (RG), blue (BG), and emission line (ELG) galaxies in the cluster. First, we explain the method of analysis to retrieve the stellar population properties of the galaxies based on the \js{} fits. Then, we describe the methods to identify ELGs. 


\subsection{\js{} fits}
\label{sec:method}

We use \baysea\ (de Amorim et al., in prep.), a parametric SED fitting code, to obtain the stellar population properties from \js.
\baysea \ is an adaptation of the method developed by \cite{lopez-fernandez2018} in order to use \jp \ magnitudes as input.
The code generates synthetic \js\ from parametric SFH models. For a given observed \js\ we perform a Markov Chain Monte Carlo (MCMC) exploration of the parameter space, thus obtaining a sample of parameters that approximates the probability density function (PDF) of the model. In this work we assumed a delayed-$\tau$ model given by

\begin{equation}
    \psi(t)=\frac {M_{ini}} {\tau^2 \left [ 1- e^{-\frac{t_0}{\tau}} \left ( \frac{t_0}{\tau}+ 1 \right )   \right ]}(t_0-t)e^{-\frac{t_0-t}{\tau}},
    \label{eq:SFH}
\end{equation}
where $t$ is the look-back time, $t_0$ is the (look-back) time when the star formation began, $\tau$ is a measurement of how extended in time the star formation was and  ${M_{ini}}$ is the total mass of formed stars.
This model also includes stellar metallicity ($Z$) and dust attenuation ($A_V$), which combined with stellar population model spectra and a foreground dust screen extinction curve, results in a model \js \footnote{The stellar population spectra are preprocessed and converted to observed-frame magnitudes for a grid of redshifts, using \jp \ filter curves. }.
The complete set of parameters is $(t_0, \tau, A_V, Z)$.
We also obtain the stellar mass ($M_\star$) from the scaling factor of the model with relation to the observed \js.
From these parameters, we calculate the mass-weighted and light-weighted ages, and rest-frame colours.

We let $100$ chains walk the parameter space for 2200 steps.
The autocorrelation time\footnote{The assessment of autocorrelation time and convergence of the chains was performed in a small sample of \js. This is a fairly manual process, as with any MCMC convergence study. We consider the burn-in phase to be over at around $5 \times$ the autocorrelation time, which we assume is conservative enough.} of the chains for this model is around 120 steps, we discard the first 1200 steps as a burn-in phase.
In the end, we get a total of 100,000 samples of the parameter space.
For each galaxy, we take the mean and standard deviation of the parameters and properties of the samples as an estimate of their expected value and uncertainties.

Emission lines are not included in the models. Because some of the NBs can be affected by strong contributions from the Balmer (H$\alpha$, H$\beta$) and optical collisionally-excited ([OIII]$\lambda$5007, 4959, [OII]$\lambda$3727, [NII]$\lambda$6589, 6548) emission lines, we remove from the fits the bands where these lines could be  at the redshift of each galaxy. In this way, we ensure that the fit is done only over the stellar continuum since the nebular continuum is negligible in most of the star-forming galaxies. Only objects with HII regions with very extreme emission lines (e.g. \ewha{} > 1000 \AA  \, \citet{gonzalezdelgado1994}) will be affected by this assumption. These galaxies are not present in this sample.

A more detailed explanation of the method with a global study of the galaxies in the AEGIS field can be found in \cite{Rosa2021}. Since the data used in this paper are a subsample from \cite{Rosa2021}, the models are computed using the initial mass function (IMF) by \cite{Chabrier2003} and the latest versions of the \cite{C&B2003} stellar population synthesis models \citep{Plat2019}. We chose the attenuation law by \citet{calzetti2000} which we added as a foreground screen. We also note that, unless stated otherwise, the term \textit{mass} refers to the stellar mass (M$_\star$) derived by \baysea.

In Fig. \ref{fig:jspectra} we show the \js \ of six galaxies (three red and three blue) along with the fits obtained with \baysea. These galaxies are identified with red and blue circles in the top panel. This serves as an example of the aspect of \jp \ data, the effectiveness of \baysea \ and it also manifests the capability of \jp \ to detect line emission (which will be exploited in Sect.\ref{sec:ELG}).

\subsection{Identification of red and blue galaxies in the cluster}

\label{sec:Masscolour}

Throughout this paper we use two different colours: $(u-r)_{\mathrm{res}}$ and $(u-r)_{\mathrm{int}}$. Both of them are rest frame colours derived from the star formation history obtained from the SED fitting (see Sect. \ref{sec:method}), but the first one is not corrected from extinction while the second one is calculated including the reddening correction in the synthetic SED.

The bimodal distribution of the AEGIS galaxy population is shown in the galaxy stellar mass-colour diagram \citep{Rosa2021}. In this work we show that $(u-r)_{\mathrm{int}}$ is more useful than $(u-r)$ to discriminate between the red star-forming and quiescent galaxies because it accounts for the fraction of red star-forming galaxy population of the sample. We use an adaptation of the criterion given by \cite{Luis2019}, previously been used by \cite{Rosa2021} to segregate the whole galaxy populations in miniJPAS in red and blue galaxies. We consider that galaxies are red if:
\begin{equation}
    (u-r)_{\mathrm{int}}> 0.16 (\log( M_{\star})-10)-0.254(z-0.1)+ 1.792
    \label{eq:red-blue}
\end{equation}
and blue otherwise.

\subsection{ELG identification}

\begin{figure}
    \centering
    \includegraphics[width=0.45\textwidth]{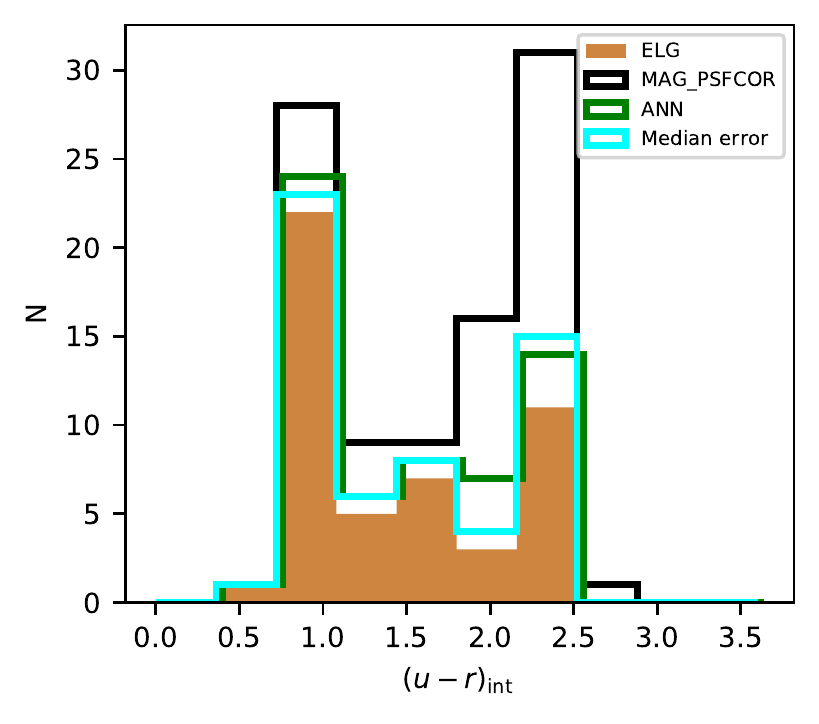}
    \caption{$(u-r)_{\mathrm{int}}$ histogram comparing the emission line galaxies selection criteria. The black line shows the distribution of all the galaxies in the AMICO catalogue. Cyan histogram shows the distribution for the median error. Same for green but with the ANN method. Orange solid histogram shows the distribution for the common galaxies for both methods. The different distributions have been shifted a little bit in the plot to show more clearly the similarities and differences between them.  }
    \label{fig:method_comparisson}
\end{figure}

\label{sec:ELG}
As seen in Fig. \ref{fig:jspectra}, the \js \ are capable of showing the line emission as excess flux in a given filter. In this section, we describe two methods to classify the galaxies as emitters or non-emitters, attending only to H$\alpha$ emission. Then, we apply the methods to the cluster catalogue to characterise the  emission line galaxy populations.

\subsubsection{Median error method}
This method is based on our \jp{} but also uses a prior based on the result from our SED fitting code, since we distinguish between red and blue galaxies. The base idea is that, when looking at the filter that is sensible to the observed line wavelength, a galaxy that presents H$\alpha$ (or H$\alpha$ +[NII], since in this method we can not separate the emission of both lines.)  emission will show a lower magnitude value in the observed \js \ ($\mathrm{m}_{\mathrm{obs}}$) than in the stellar continuum fit ($\mathrm{m}_{\mathrm{fit}}$). However, it is not enough to simply consider that $\mathrm{m}_{\mathrm{fit}}-\mathrm{m}_{\mathrm{obs}}>0$. We must establish a threshold value. A first consideration to make is that the difference must be greater than the observed error. In order to estimate the observed error of the fitted stellar continuum, we consider the median error in the 5 filters closer to the band where H$\alpha$ is, symmetrically distributed. If we only choose 3 filters, our estimation could be contaminated with other lines close to H$\alpha$, such as [SII]. Choosing 7 filters only changes the final set in one galaxy. Choosing more filters would mean estimating the continuum too far from H$\alpha$, given the width of \jp \ filters. Besides, blue galaxies in our sample are noisier than red galaxies. The median S/N in the five closest filters to H$\alpha$ at the cluster redshift is almost three times better for red galaxies than for blue galaxies. We find that the larger magnitude difference of the blue galaxies is not enough to compensate their worse S/N. This implies that if the same threshold (three sigma) is applied, there will be a bias towards the detection of less blue emission line galaxies. Lastly, we must take into account that due to the uncertainties in the \photozbest{} determination, the filter with the closest central wavelength to the calculated line wavelength might not be the one that is showing the line emission. This is a consequence of our method being fine tuned to fit a set of galaxies observed with spectroscopy in the cluster area (see Appendix~\ref{appendix:GMOS}). Therefore, our method proceeds as follows. First, we find the closest filter to the observed line wavelength. Then, we calculate the median error of that filter and the four adjacent ones (symmetrically distributed) $\eta ( \epsilon_{\mathrm{m}_{\mathrm{obs}}})$. Finally, we look at the closest filter, the previous and the next one, and we classify the galaxy as an emission line galaxy if one of those filters satisfies that:
\begin{equation}
    \mathrm{m}_{\mathrm{fit}}-\mathrm{m}_{\mathrm{obs}}> \theta \cdot \eta (\epsilon_{\mathrm{m}_{\mathrm{obs}}})
\end{equation}
\noindent where $\theta=1$ for blue galaxies and $\theta=3$ for red galaxies in order to account for their better S/N in the filters closer to the H$\alpha$ wavelength at the cluster redshift. 
These values of $\theta$ were chosen in order to account for this differences in the S/N ratio. They have been tested with the data from GMOS spectroscopic observations of 13 galaxies with clear H$\alpha$ emission in the spectra.

This method allows us to identify emission line galaxies in the cluster, but we do not use it to estimate the fluxes of the lines. ANN method is more useful at this respect.  

\subsubsection{ANN}
This method uses the Equivalent Width of H$\alpha$, EW(H$\alpha$) predictions made by \cite{Gines2021} ANN. In that work, two different ANNs are trained using synthetic photometry (in the sense that real spectra are processed to obtain \jp \ magnitudes) obtained from the Calar Alto Legacy Integral Field Area survey \citep[CALIFA,][]{CALIFA2012} and the Mapping nearby Galaxies at Apache Point Observatory survey \citep[MANGA,][]{MANGA2015}. One of the ANN is trained to calculate the EW of H$\alpha$, H$\beta$, [OIII] and [NII]. The other ANN is trained in order to classify the galaxies into ELG and quiescent galaxies. 
CALIFA and MANGA galaxies contain millions of spaxels with different astrophysical conditions, which include regions with high and low star-formation activity, variations in the gas-phase metallicity or the dust distribution. Furthermore, CALIFA and MANGA survey contain galaxies in different environments (clusters, groups and field) since they were selected to avoid environmental bias \citep{Walcher2014,Wake2017}. Therefore, we do not expect our prediction to be unlikely in the cluster under study. Nevertheless, by construction our train set include in average less amount of spaxels ionized by the presence of AGN or shocks waves. In \cite{Gines2021} we showed that we do not miss a fraction of AGN larger than $3$~\% over the whole sample of galaxies used from SDSS. This emphasises that the transfer from the training sample to our current data is trustworthy.

As explained in \cite{Gines2021}, there is a minimum measurable EW for a photometric filter. Therefore, the criteria we use is simply to consider the galaxy as an emission line galaxy if the EW given by the ANN is greater than the minimum measurable EW, taking the error bars into account. This is; 
\begin{equation*}
    \frac{\Delta '}{\mathrm{S/N}-1} <  \mathrm{EW_{H\alpha_{\mathrm{ANN}}}}+ \epsilon \mathrm{EW_{H\alpha_{\mathrm{ANN}}}} \ ; \  \mathrm{EW_{H\alpha_{\mathrm{ANN}}}} > \epsilon \mathrm{EW_{H\alpha_{\mathrm{ANN}}}} 
\end{equation*}
\noindent where $\Delta '$ is the equivalent width of the filter and can be calculated as 

\begin{equation}
    \Delta ' = \frac{\int \lambda T (\lambda ) \mathrm{d}\lambda}{\lambda_z T (\lambda_z )} 
\end{equation}

\noindent where $T$ is the normalised transmittance of the filter and $\lambda_z$ is the observed emission line wavelength. To compute the minimum measurable EW we find the J-PAS filter whose central wavelength is the closest one to the observed H$\alpha$ wavelength given the galaxy's redshift. Since the minimum EW is associated to a certain filter, the $\Delta '$ and S/N parameters must be considered in the same filter and it does not make any sense to consider a median value for any of them. Also, since we are using the ANN predictions, here we can separate the H$\alpha$ emission fom the [NII] emission.

This method is very useful not only to identify emission line galaxies, but also to predict the EW of H$\alpha$, H$\beta$, [NII], and [OIII] lines and their respective ratios ([NII]/H$\alpha$, [OIII]/H$\beta$). We are able to reach a precision in the $\log$([NII]/H$\alpha$) of $0.09$~dex for SF galaxies and average S/N $\sim$10 in the \js{}. This is independent of the redshift of the galaxies as we prove in \citep{Gines2021} where we tested our results with a sample of SDSS galaxies within the redhisft range of $0 < z < 0.35$.
 
\subsubsection{ELG final set}
We now apply both methods to all the galaxies in the cluster. The median error method selects 57 galaxies as galaxies with emission lines, while the ANN method selects 50 galaxies in total. We decide to be more conservative and consider as ELG population the intersection of both groups, since this defines a more robust subset. A total of 49 galaxies remains. A comparison of the three sets with the whole cluster colour distribution can be seen in Fig. \ref{fig:method_comparisson}. The $(u-r)_\mathrm{int}$ distribution is very similar in both methods, peaking around $(u-r)_\mathrm{int}\approx 1$ and with a lower peak at $(u-r)_\mathrm{int}\approx 2.2$. This peak is easily understood when looking at the general distribution, since there is also a peak at this value, even greater than the bluer one. When defining the median error method and establishing the values of the multipliers there was a risk to create a greater bias towards blue galaxies greater than desired in order to account for the larger errors. Comparing its histogram with the ANN, one we can see that the proportion of blue and red galaxies remains very similar, so we can trust this method. As a final comment, these results are coherent with what we would expect, since most of the ELG are blue.

\section{Characterisation of the galaxy populations}
\label{sec:charact}

\begin{table*}[]
    \centering
     \caption{Mean and standard deviation values of the stellar population properties of the galaxies in the cluster. The properties are for red and blue galaxies (RG, BG), emission lines galaxies (ELG), ELG with red (ELG-R) or blue (ELG-B) colours, star forming galaxies (SF), and  galaxies with an AGN.    }
    \begin{tabular}{c c c c c c c c c}
    \hline
    \hline
         {\small Property} & {\small Galaxies} & {\small RG}  & {\small BG}  & {\small ELG} &  {\small ELG-R} &  {\small ELG-B} &  {\small SF} &  {\small AGN}\\
         \hline
$\log M_\star$ & $10.0 \pm 0.65$ & $10.4 \pm 0.32$ & $9.63 \pm 0.64$ & $9.89 \pm 0.71$ & $10.5 \pm 0.36$ & $9.62 \pm 0.65$ & $9.56 \pm 0.60$ & $10.5 \pm 0.42$\\
$ A_V$ & $0.57 \pm 0.43$ & $0.32 \pm 0.19$ & $0.84 \pm 0.44$ & $0.63 \pm 0.42$ & $0.32 \pm 0.25$ & $0.76 \pm 0.40$ & $0.65 \pm 0.41$ & $0.49 \pm 0.35$\\
$ \log Z_\star>$ & $0.09 \pm 0.45$ & $0.29 \pm 0.20$ & $-0.1 \pm 0.53$ & $0.05 \pm 0.50$ & $0.35 \pm 0.19$ & $-0.0 \pm 0.54$ & $-0.0 \pm 0.53$ & $0.35 \pm 0.25$\\
$ (u-r)_\mathrm{res}$ & $2.04 \pm 0.51$ & $2.42 \pm 0.10$ & $1.64 \pm 0.48$ & $1.85 \pm 0.55$ & $2.43 \pm 0.08$ & $1.59 \pm 0.47$ & $1.60 \pm 0.50$ & $2.31 \pm 0.28$\\
$ (u-r)_\mathrm{int}$ & $1.67 \pm 0.60$ & $2.21 \pm 0.16$ & $1.10 \pm 0.30$ & $1.44 \pm 0.58$ & $2.22 \pm 0.16$ & $1.09 \pm 0.30$ & $1.18 \pm 0.46$ & $1.98 \pm 0.43$\\
$ <\log age>_\mathrm{M}$ & $9.44 \pm 0.24$ & $9.61 \pm 0.13$ & $9.26 \pm 0.21$ & $9.34 \pm 0.23$ & $9.53 \pm 0.17$ & $9.26 \pm 0.21$ & $9.27 \pm 0.23$ & $9.47 \pm 0.22$\\
$\tau / t_0 $ & $0.52 \pm 0.60$ & $0.12 \pm 0.02$ & $0.94 \pm 0.63$ & $0.71 \pm 0.66$ & $0.11 \pm 0.02$ & $0.97 \pm 0.64$ & $0.95 \pm 0.68$ & $0.23 \pm 0.27$\\
         \hline
    \end{tabular}
   
    \label{tab:sp_auto_psfcor}
\end{table*}

In this section, we analyse the stellar population properties of the galaxies belonging to the cluster. First, the sample is divided in red and blue galaxies. Then, ELG are characterised by their stellar populations by dividing also the galaxies in star-forming (SF) and galaxies with an active galactic nucleae (AGN). The average and dispersion values of the stellar population properties are summarised in Table~\ref{tab:sp_auto_psfcor}.

\subsection{The red and blue galaxy populations}

\begin{figure*}
\centering
\includegraphics[width=\textwidth]{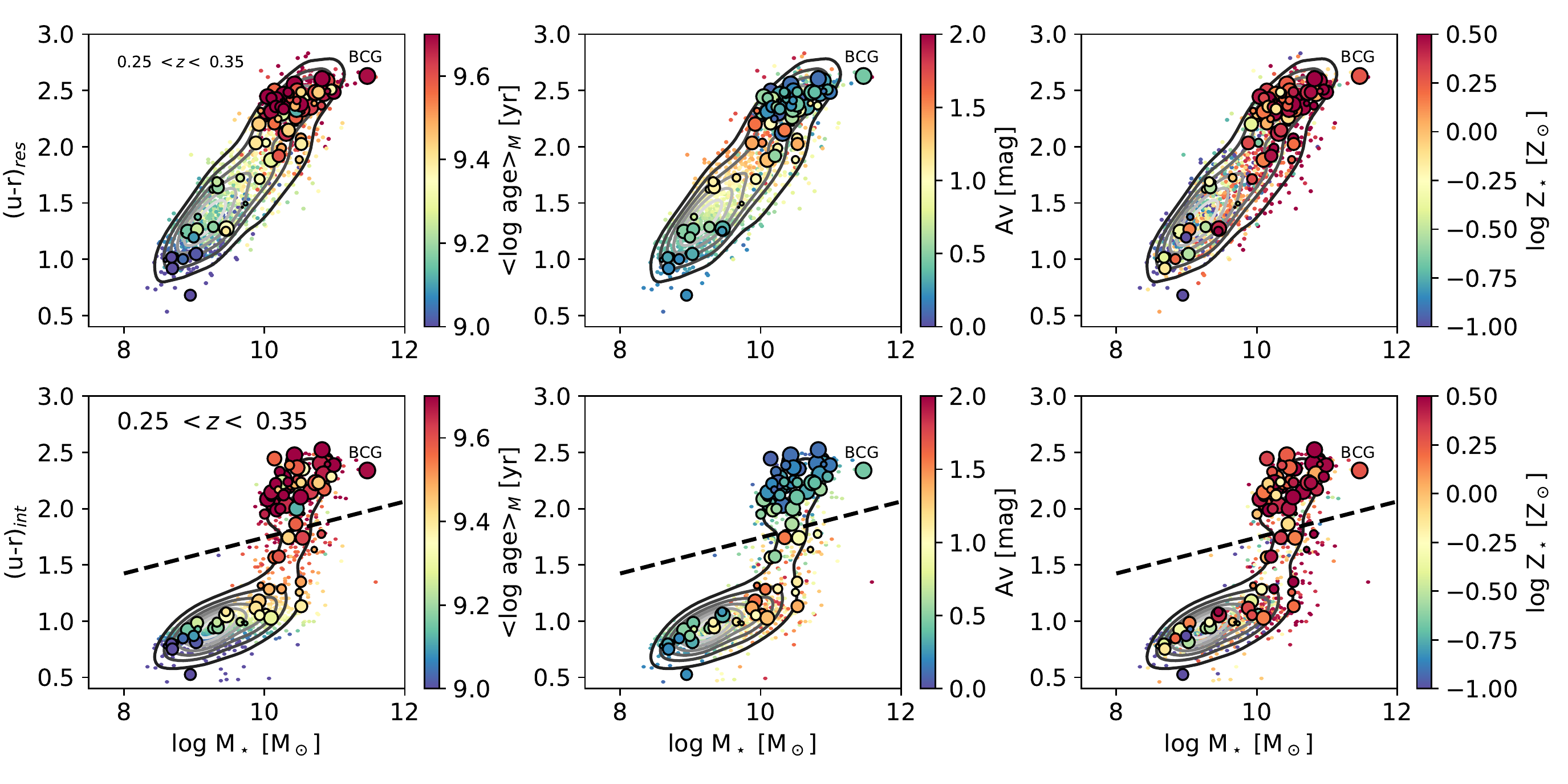}
\caption{$(u-r)_{\mathrm{res}}$ (top panels) and $(u-r)_{\mathrm{int}}$ (bottom panels) colour vs stellar mass for the  redshift bin 0.25$< z <$0.35 derived by \baysea \ from the AEGIS galaxy populations (contour) and galaxy cluster members (circles). The coloured bar shows the distribution of the stellar population properties age, extinction and metallicity (from left to right). The size of the circles indicates the probability of the galaxy to be member of the cluster. The position of the brightest galaxy in the cluster (BCG) in each panel is marked. The dashed line in the $(u-r)_{\mathrm{int}}$ divides blue galaxies (below the line) and red galaxies (above the line).
}
\label{fig:masscolour}
\end{figure*}

\begin{figure*}
\centering
\includegraphics[width=\textwidth]{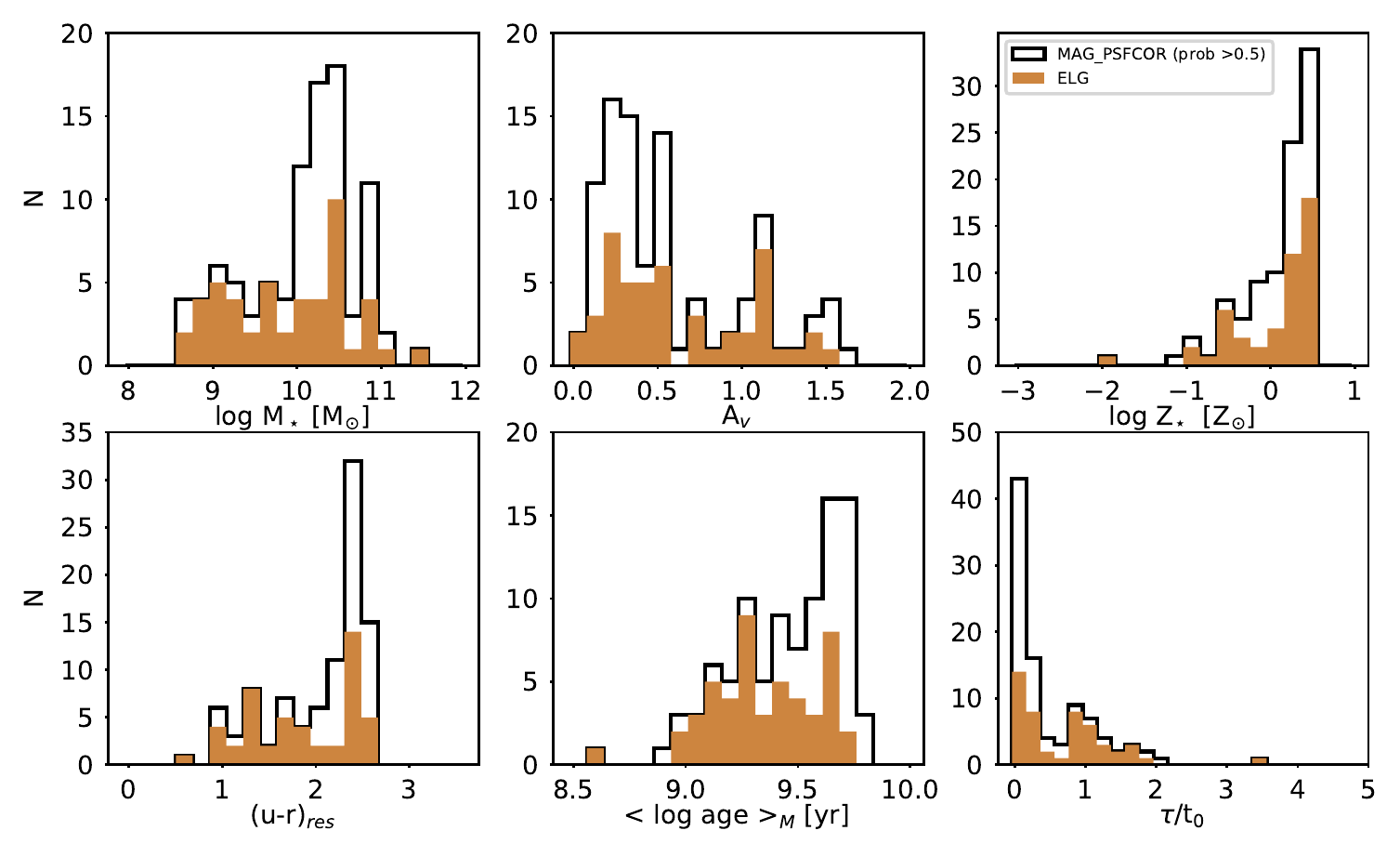}
\caption{Stellar population properties distribution for the emission line galaxy population. Black histogram shows the distribution for all the galaxy members in the \magpsfcor \ photometry. Brown solid histogram shows the distribution for the objects selected as emission line galaxies. From left to right and from upper to bottom: stellar mass, extinction, stellar metallicity, $(u-r)_{\mathrm{res}}$ colour, mean mass-weighted age, and ratio between the SFR parameters $\tau$ and t0. 
}
\label{fig:em_line_hist}
\end{figure*}

\begin{figure}
    \centering
    \includegraphics[width=0.49\textwidth]{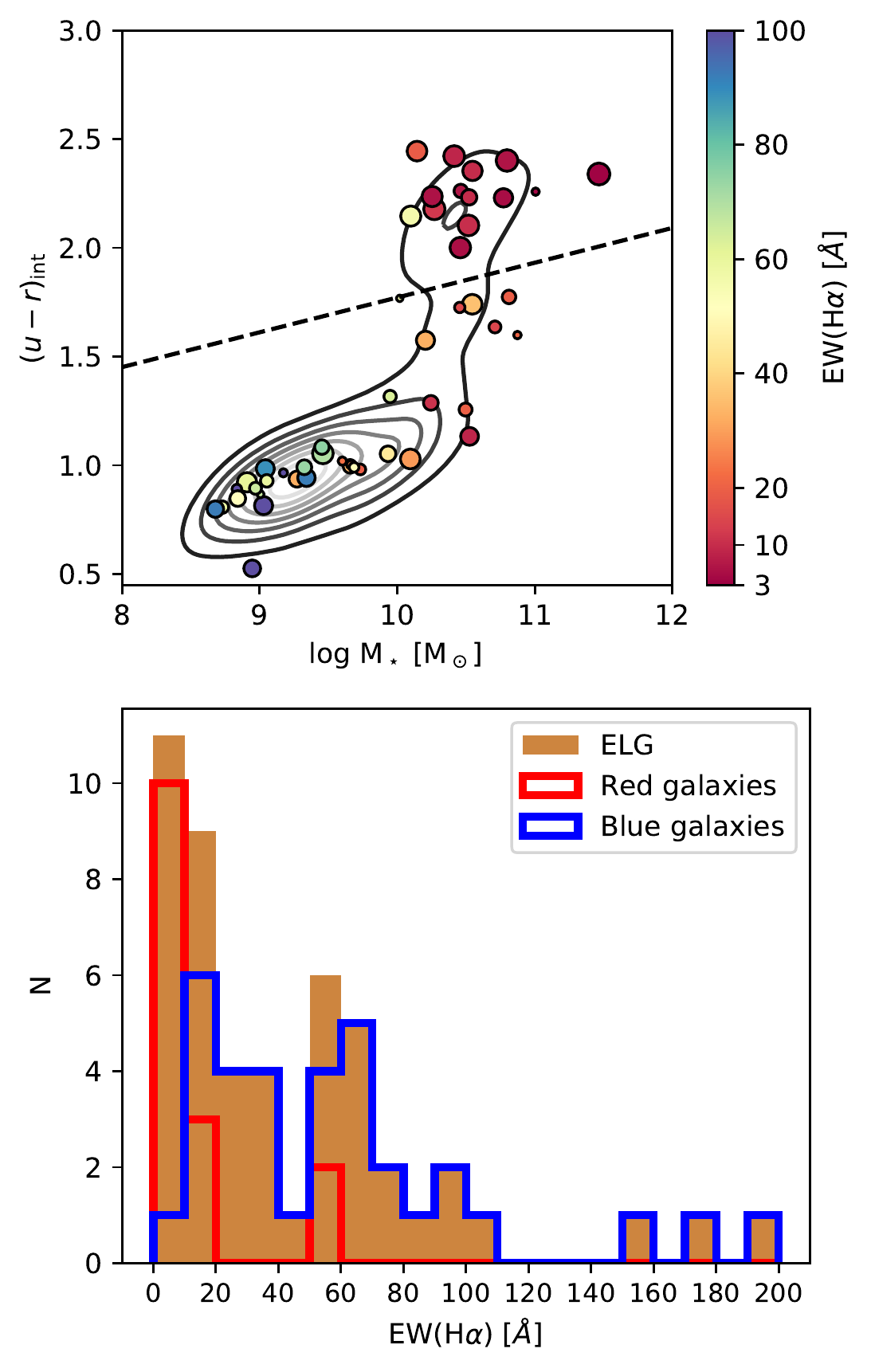}
    \caption{Top panel: colour-mass diagram for the ELG galaxy population. The colour bar shows the H$\alpha$ equivalent width. Galaxies above the black dashed line are considered to be red. Galaxies below are considered to be blue. Bottom panel: H$\alpha$ equivalent width distribution histogram. Green histogram shows the distribution of the galaxies selected with the ANN method. Orange solid histogram shows the distribution for the selected galaxies. Red histogram shows the distribution of the selected red galaxies. Same holds true for blue histogram and blue galaxies.}
    \label{fig:EW_Ha}
\end{figure}

The bimodal distribution of the red and blue galaxies in the cluster can be seen in the colour-mass diagram (see Fig.~\ref{fig:masscolour}) The comparison between the $ (u-r)_\mathrm{res}$ and the $(u-r)_\mathrm{int}$ shows how the extinction correction moves a significant number of galaxies from the redder (upper) regions of the diagram to the bluer regions (below the black dashed line). The colour code also shows how, on average, redder galaxies are older and more metal rich. Galaxies with larger extinctions are located in the middle region of the diagram, which could be considered as an equivalent of the green valley. The comparison with \cite{Rosa2021} results for the whole AEGIS catalogue shows that the distribution of these properties in the colour-mass diagrams remains the same. This would indicate that, for fixed values of the colour and mass, the effect of the environment on these properties is negligible.
We find that, on average, red galaxies are more massive than blue galaxies by $\sim 0.8$~dex. Blue galaxies also show a larger variance in mass. The extinction $A_V$ is significantly larger on average ($\sim 0.5$~mag) for blue galaxies. It is expected because most of the blue galaxies are star forming, and the extinction that young stars experience is almost  double than that for the old stellar population \citep{charlot2000}. In contrast, blue galaxies are less metal rich than red galaxies by $\sim 0.1$~dex.  On average, red galaxies are older by $0.4$~dex. The value of  $\tau/t_0$  is nine times larger for blue galaxies than for red galaxies; thus, the star formation lasts longer in the blue galaxy population. 

The total fraction of red galaxies is $0.52$ ($0.48$ for blue galaxies). The fraction of red galaxies in the whole catalogue of \mjp \ at the cluster's redshift, obtained by \cite{Rosa2021},  using \baysea, is around $0.2$ or even lower. This is supported by works in the literature such as \cite{Balogh2004}. If we assume a symmetric distribution within $R_{200}$, and that all the missing galaxies there were blue (worst case scenario), the fraction of red galaxies inside $R_{200}$ would be 0.55 (compared to the current observed fraction of $0.62$ inside $R_{200}$), which is still higher than the fraction of red galaxies in the field. Instead, if we assumed a symmetrical distribution keeping the same amount of blue and red galaxies in the missing area, we would find a fraction of red galaxies even larger ($0.64$).

\subsection{ELG population}

In Fig.~\ref{fig:em_line_hist} we compare the distribution of the stellar properties of the ELG with the whole sample, and we summarise them in Table~\ref{tab:sp_auto_psfcor}. We find that their values span the same ranges than the properties of the whole catalogue. However, the distribution themselves are different. The stellar mass still peaks at $\log M_\star \approx 10.5$~[$M_\odot$], but the contribution of galaxies with $\log M_\star < 10$~[$M_\odot$] becomes more significant. In fact, most of the galaxies in such range are classified as ELG. On average, ELG are less massive than the whole sample by 0.1 dex.
The distribution of $A_V$ shows that most of the galaxies that are not selected as ELG show values lower than 0.5, but the distribution remains similar (the average only becomes $0.06$~mag lower). A similar behaviour is found for the metallicity, where most of the galaxies with $ \log Z_\star\lesssim$-0.5 are ELG, but the peak of the distribution is still the same as the whole set. The average only becomes lower by $0.04$~dex.
Nonetheless, the distribution of $ (u-r)_\mathrm{res}$ changes significantly. The peak of the distribution is still found at $ (u-r)_\mathrm{res}\approx 2.5$~mag, but most of the galaxies with $ (u-r)_\mathrm{res}< 2$ are ELG, and only a few galaxies with $ (u-r)_\mathrm{res}> 2$ are ELG. Moreover, the peak of the stellar ages is now found at  $ <\log age>_\mathrm{M} \approx 9.25$, with most of the young galaxies being ELG and only a few of the old galaxies showing emission lines (ELG are younger by $0.1$~dex on average). Furthermore, most of the galaxies  with $\tau /t_0 \lesssim 0.8$ are not ELG, and almost all the galaxies $\tau /t_0 \gtrsim 0.8$ are ELG, and the average value of ELG is almost a 50~\% larger than the average of the whole sample.

These results show that ELG have properties similar to the BG population. However, they show differences that suggest that ELG is a mix of red and blue populations, where RG are significantly less abundant than the BG population. To investigate further this point we explore the colour-mass diagram and the distribution of the inferred EW(H$\alpha$), dividing the ELG into red (ELG-R) and blue (ELG-B) galaxies (Fig. \ref{fig:EW_Ha}). We see that galaxies with the lowest predicted EW (<10 \AA \ approximately) are all red galaxies, while all galaxies  above this value are all blue except for three of them. Two of them are particularly notable, having a predicted EW above 50 \AA. However, when looking at the spectra we find that the inferred emission may be a result from incomplete background subtraction due to fringing effect that suffer some of the red filters, that is translated into a variation of the measured magnitude that could be interpreted as an emission line by the ANN, due to the magnitude difference among one filter and its adjacent ones. With that exception, we can conclude that ELG-R are characterised by low estimated values of EW(H$\alpha$), while ELG-B have EW(H$\alpha$) $>$ 10 \AA. Taking into account that our ELG-R are more massive than our ELG-B, we find that our results are coherent with the EW(H$\alpha$)-mass relation found in the literature \citep[see, e. g.][]{Fumagalli2012, Sobral2014, Khostovan2021}.

\subsection{Star-forming galaxies and AGN populations}
\label{sec:AGN}

\begin{figure*}
    \centering
    \includegraphics[width=\textwidth]{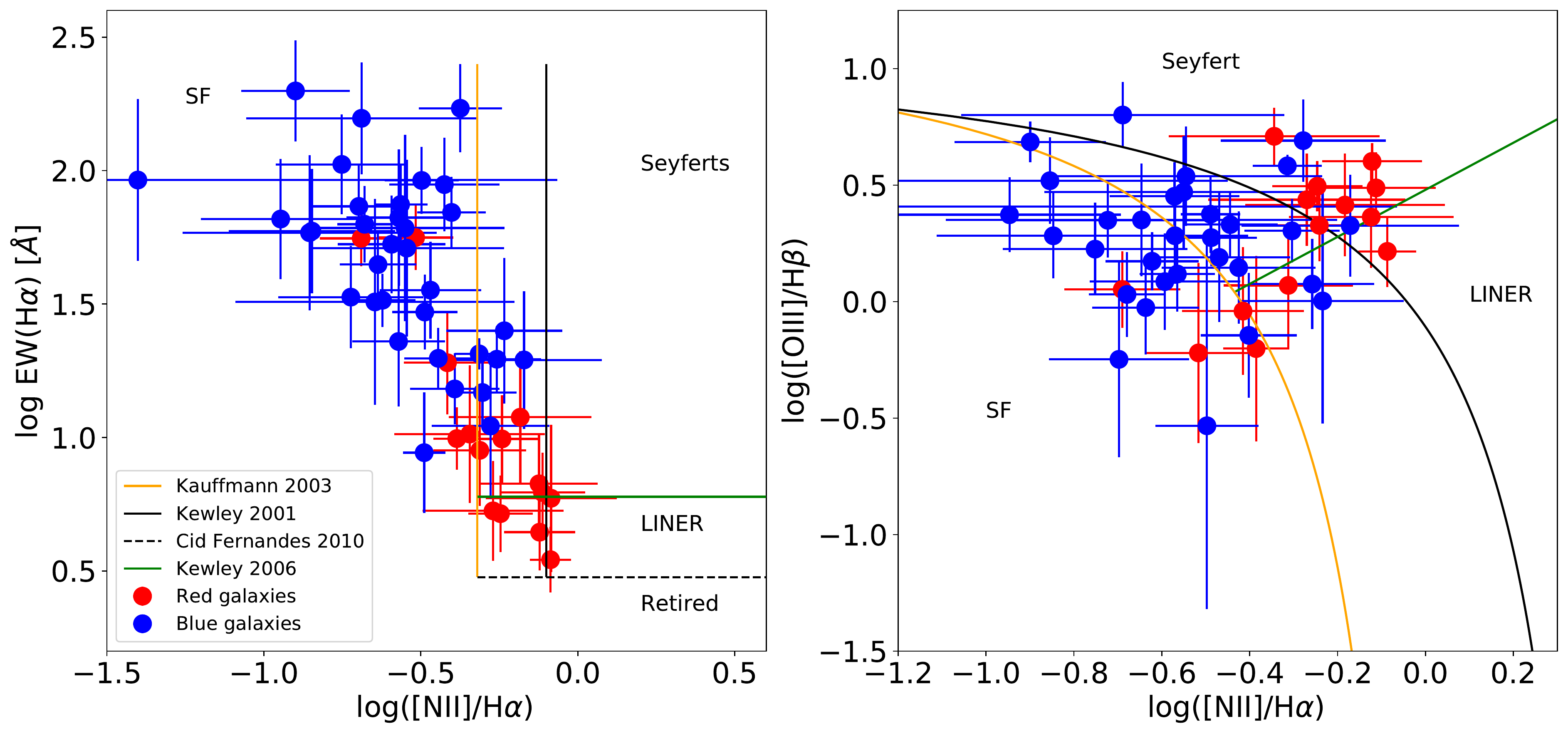}
    \caption{Left panel: WHAN diagram with the galaxies classified as emission line galaxies. Red points represent red galaxies and blue points, blue galaxies. The solid orange and black lines represent the \cite{WHAN_1} transposition of the \cite{Kauffmann2003} and \cite{Kewley2001} SF-AGN distiction criteria, and the green solid line represents the transposition of the \cite{Kewley2006} made by \cite{WHAN_1}. The dashed black line represents the distinction between retired galaxies and LINERs \citep{WHAN_2}. Right panel: BPT diagram for the emission line galaxy population. The colour coding is the same as the left panel.}
    \label{fig:WHAN+BPT}
\end{figure*}

The ELG population can be a mix of star-forming galaxies (SF) and AGN galaxies. To find the abundance of these two classes,  we use two different diagrams: the WHAN diagram, introduced by \cite{WHAN_1} (CF10), and the BPT diagram \citep{BPT}. There are several works that present their own criteria to separate the SF, Seyferts and LINERs in the BPT diagram, but we stick to three of them: we use the results from \cite{Kauffmann2003} (K03) and \cite{Kewley2001} (K01) to distinct SF galaxies from galaxies with a potential AGN, and we use the transposition to this diagram made by \cite{WHAN_1} of the separation criteria between Seyferts and LINERs found by \cite{Kewley2006} (K06). In the case of the WHAN diagram, we use the criteria from  CF10 and \cite{WHAN_2} (CF11). In this work, several criteria are presented, but for consistency we choose the transpositions made in these works (CF10 and CF11) of the same criteria used in the BPTv(K03 and K01) in addition the criteria of those articles where galaxies with EW(H$\alpha$)<3\AA  \ are considered as retired galaxies \footnote{This is the name given by CF10 to red-quiescent galaxies with weak H$\alpha$ emission, that is probably produced by post-AGB stars.}.  
An example of each type of galaxy, with its \js \ and its position in the WHAN and BPT diagrams (respectively) can be seen in Fig \ref{fig:Examples_AGN}. This figure is useful to explain more clearly how we interpret the position of galaxies in these diagrams.

Figure~\ref{fig:WHAN+BPT} shows the WHAN and BPT diagrams with all the ELG population. Table \ref{tab:AGN} shows the classification for each of these galaxies in both diagrams.
Since a different classification can be derived from each diagram, and due to the error bars obtained for many galaxies, it is not trivial to assign a label to each ELG. Therefore, we use a probabilistic approach in the following way: we calculate the area of the error box in the WHAN diagram, and we calculate the fraction of this area that falls in each of the diagram regions. We define this fraction as the probability representing how likely it is for that galaxy to be a SF, Seyfert, LINER, retired or composite (SF-Seyfert or SF-LINER) galaxy. The error bars plotted in this figure take into account the correlation among the emission lines through the calculation of the Pearson coefficient and its inclusion in the error budget.

We find that 32 galaxies ($65.3$~\% of the ELG) have a probability larger than $0.7$ associated to the SF region (33 above $0.5$, representing $67.3$~\%). We select these galaxies as the SF population. Only one galaxy (which represents 2~\% of the ELG) has a combined probability in the Seyfert and LINER region larger than 0.5 in the WHAN diagram. The rest of the galaxies are difficult to uniquely identify as AGN using only the WHAN diagram. Due to this,  we select as AGNs the galaxies that are above the K01 curve in the BPT diagram if the probability to be SF in the WHAN diagram is less than 0.5. With these criteria, only 2470-3670 needs to be excluded from the AGN sample (see Appendix~\ref{sec:ELG_class} and Fig \ref{fig:Examples_AGN}). Galaxies between the K01 and K03 lines likely have contributions from AGN, but we cannot resolve whether they are Syferts, LINERs or composite galaxies. Thus, we do not include them as part of the AGN sample, neither as SF if they are not classified as SF in the WHAN diagram. 

In Table~\ref{tab:sp_auto_psfcor} we summarise the stellar population properties of these galaxies. If we compare the SF galaxies with the blue ELG we find that the differences in the average and standard deviation values are negligible except for the extinction $A_V$, which are lower for SF galaxies, and $(u-r)_{int}$ colours, which are slightly redder (but with a greater standard deviation) for SF galaxies. This indicates that most of the blue ELG are SF galaxies.

The comparison between the values of the red ELG and AGN populations shows that the main difference between them resides in a larger extinction on average for the AGNs, slightly bluer $(u-r)_{res}$ and $(u-r)_{int}$ colours, younger ages and higher values of $\tau/t_0$. This indicates that the sample selected as AGN is a mixture of blue and red ELGs, and that we are not able to fully separate the contributions of pure AGNs from star formation in galaxies, or that most of these galaxies are actually composite. 

We have focused on the H$\alpha$ emission in order to select our ELG sample. This line  can be used as a tracer of the star formation \citep[see e.g.][]{Kennicutt1998, Kennicutt2012, Kewley2002, Garn2010, Oteo201, CT2015} or the presence of AGNs \ \citep[see e.g.][]{Osterbrock1985,Veilleux1987,Osterbrock1989, Kewley2001, WHAN_2}.  Therefore, taking into account the relation between galaxies in the blue cloud and a higher star formation than galaxies in the red sequence, \citep[see e.g.][]{Kauffmann2003a, Kauffmann2003b, Baldry2004, Brinchmann2004, Gallazzi2005, Mateus2006, Mateus2007}, it is reasonable that an important fraction of our selected ELG are blue galaxies that are generally classified as star-forming galaxies according to the WHAN and BPT diagrams, and that red galaxies are generally classified as LINERs or retired galaxies. A similar discussion with compatible results can be found in the works by \cite{Chies2015, RodriguezdelPino2017}. 

\subsection{The star formation rate}
\label{sec:SFR}

\begin{figure}
\centering
\includegraphics[width=0.45\textwidth]{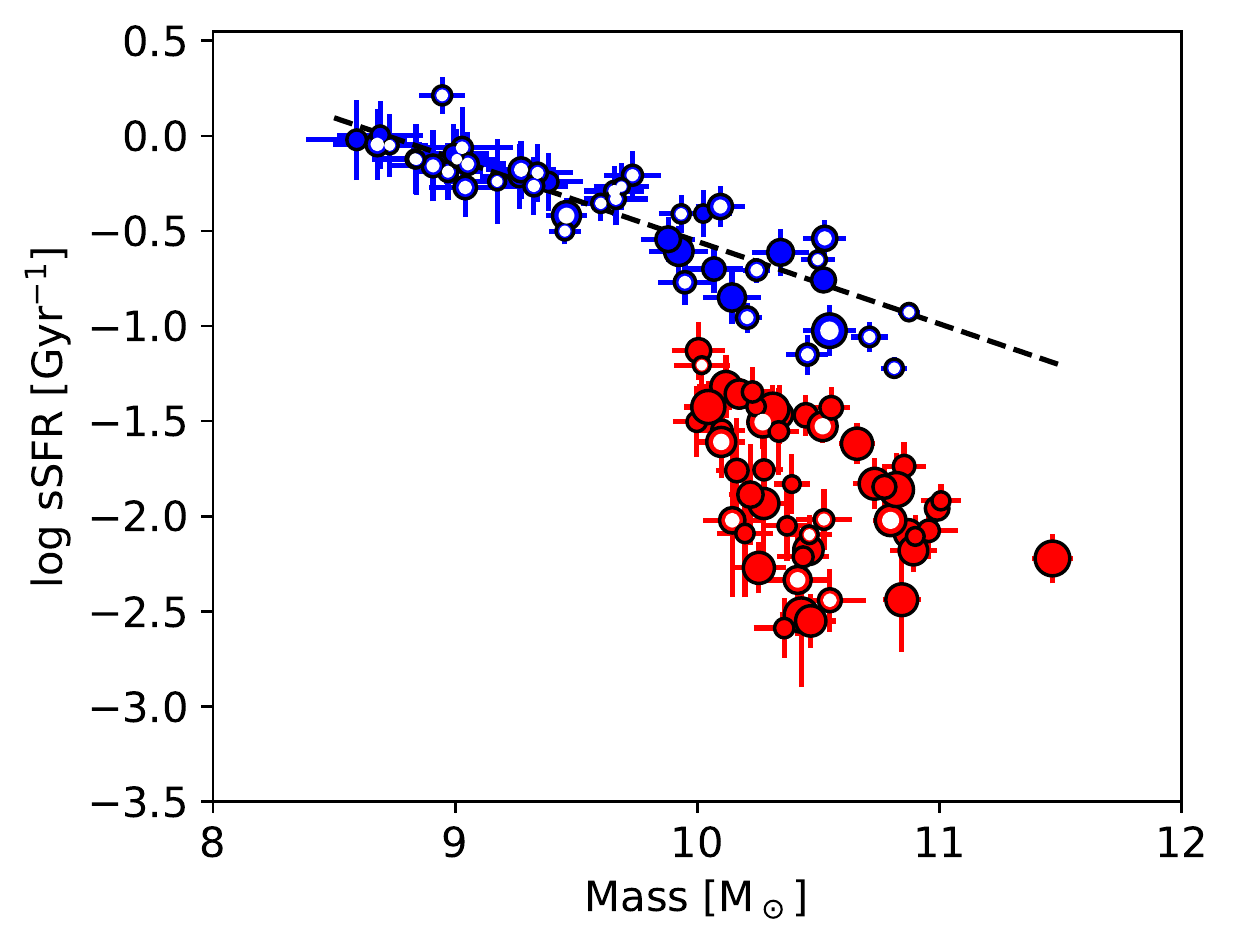}
\caption{Star formation main sequence. Red dots represent the red galaxies. Blue dots represent  blue galaxies. White dots represent the ELG population (selected in Sect.\ref{sec:ELG}) with inferred EW(H$\alpha)>6$~\AA. Dot size is proportional to the inverse distance to the BCG} 
\label{fig:main_sequence}
\end{figure}

In order to calculate the star-formation rate (SFR) and the specific SFR (sSFR = SFR/M$_\star$), we use the star formation history (SFH), derived from the SED-fits, to sum the stellar mass formed in the last 20 Myr and divide it by 20 Myr.\footnote{In this text, sSFR is expressed in units of Gyr$^{-1}$; and log sSFR is the decimal logarithm  of the sSFR}. The mean and standard deviation values we obtain for the sSFR for each set of galaxies are $0.25 \pm 0.32$~Gyr$^{-1}$ for the whole sample, $0.020 \pm 0.016$~Gyr$^{-1}$ for red galaxies, $0.49 \pm 0.32$~Gyr$^{-1}$ for blue galaxies, $0.35 \pm 0.35$~Gyr$^{-1}$ for the ELGs, $0.016 \pm 0.015$~Gyr$^{-1}$ for the red ELGs, $0.50 \pm 0.32$~Gyr$^{-1}$r for the blue ELGs, and $0.48 \pm 0.34$~Gyr$^{-1}$ for SF galaxies. 

We find that the mean value of the sSFR of the blue galaxies is $\sim 25$ times larger than the mean value of the red galaxies. This is accordance with the literature \citep[see e.g.][]{Kauffmann2003a, Kauffmann2003b, Baldry2004, Brinchmann2004, Gallazzi2005, Mateus2006, Mateus2007}. The difference in the values obtained for the red galaxies and the red ELGs is negligible, as well as the difference between blue galaxies and blue ELGs. 

The star-forming main sequence \citep{Noeske2007} is a relation between the SFR and the stellar mass of galaxies in the form of a power law \citep[see e.g.][]{Elbaz2007,Speagle2014,Sparre2015,cano2016spatially, vilella2021j}. The work by \cite{Nantais2020} supports that the relation remains constant with density, and \cite{Speagle2014} and \cite{Santini2017} works find no variation with redshift in the slope, but \cite{Noeske2007} find variations with redshift.

We study the main sequence of the star formation in Fig. \ref{fig:main_sequence}.We find that blue galaxies are well placed in the main sequence. The low mass/ high sSFR end of the main sequence is dominated by blue ELGs. This is compatible with young stars as the main mechanism of H$\alpha$ emission for blue galaxies, as seen in Sect.\ref{sec:AGN}. Meanwhile, red ELGs are mainly found in the low sSFR region, so their H$\alpha$ emission is probably due to other mechanism, such as the presence of an AGN. We fit a main sequence of the star formation using the SF galaxies. 
The obtained fit (see black dashed line in Fig. \ref{fig:main_sequence}) is: 
\begin{equation}
    \log \mathrm{sSFR}= 
    (-0.43 \pm 0.07)\log M_\star + (3.78\pm 0.64)  
\end{equation}

Translating this fit to SFR instead of sSFR, the obtained slope is $0.57$. These results are lower than the ones obtained by \cite{Gines2022} analysing the whole AEGIS field. In that work, they calculate the SFR through the H$\alpha$ flux and the SFH provided by \baysea. The values of the slope (in the SFR vs Mass fit) are both higher than our results. This means that sSFR decreases more rapidly with mass for the galaxies in this cluster than for galaxies in lower density environment, that is the dominant population in AEGIS \citep{Rosa2022}. However, the SF galaxies with 
$\log M_\star <9.8$~$M_\odot$ shows a flatter slope that  would suggest that the sSFR is almost independent of the mass. This also holds truth for the results from other works, such us \cite{boogaard2018muse}, \cite{vilella2021j}, \cite{puertas2017aperture}, \cite{renzini2015objective}, \cite{zahid2012census}, \cite{shin2021metal}, \cite{belfiore2016sdss}, \cite{cano2016spatially}, \cite{cano2019sdss}, \cite{sanchez2018sdss}.

\section{Discussion}
\label{sec:discussion}

\begin{figure*}
\centering
\includegraphics[width=\textwidth]{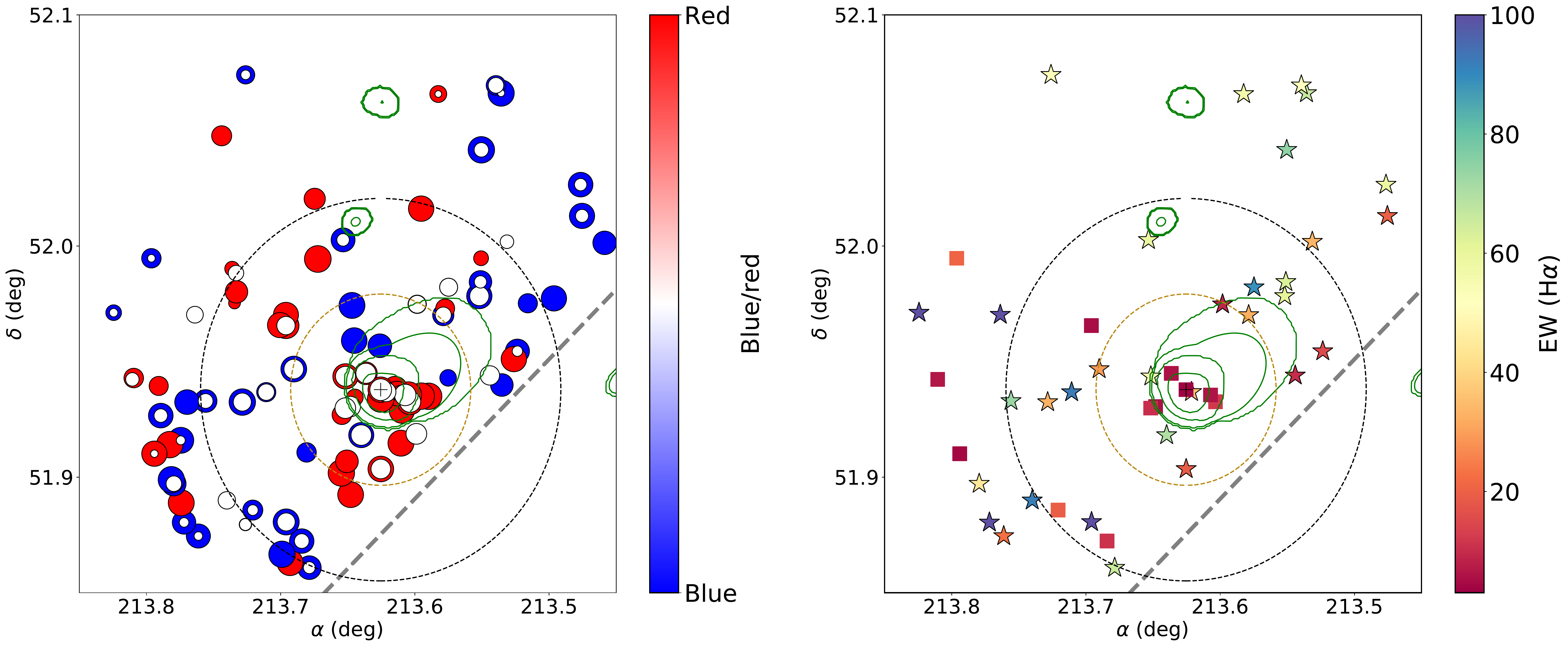}
\caption{Left panel: Spatial distribution of the red, blue, and  emission line galaxy populations. Blue dots represent blue galaxies. Red galaxies are represented with red dots. White dots over red and blue dots represent the ELG. Dot size is proportional to the AMICO association probability. The dashed golden and black circles represents the $0.5$~R$_{200}$ and R$_{200}$ distances to the BCG, respectively. The grey dashed line represents the limit of the FoV of \mjp. The black cross represents the position of the BCG. Green contours represent the X-ray emission in the 0.5-2 keV energy band from XMM data \citep{Bonoli2020}. Energy levels are $3.654 \times 10^{-16}$, $1.218 \times 10^{ -15}$, $3.654 \times 10^{ -15}$ and $1.218 \times 10^{-14}$~ergs~s$^{-1}$~cm$^{-2}$ arcmin$^{-2}$. Rigt panel: spatial distribution of the SF and AGN. Stars represent SF galaxies and squares the AGNs. The color code represent the inferred} EW(H$\alpha$). The rest of the symbols are the same as in the left panel.
\label{fig:spatial_distribution_with_contours}
\end{figure*}

The spatial distribution of the galaxy populations and the variation of galaxy properties with the cluster-centric radius is a key information to know the role of  environment for quenching the star formation in galaxies  \citep{Donnari2021,Dacunha2022, Niemiec2022}, Galaxy-galaxy or galaxy-ICM interactions are more efficient at the cluster centre where the density of galaxies and the density of the gas are higher. Therefore, it is expected that environmental processes are more efficient inside the virial radius  \citep{Alonso2012, Raj2019}. However, other processes, such as galaxy harassment, ram-pressure stripping, and starvation can act outside the virialised region, being also effective at the cluster periphery  \citep{Bahe2013, Zinger2018, Lacerna2022}. The analysis of the galaxy populations, SFR and SFH have been proven to be very useful to study quenching and cluster formation scenarios \citep[see e.g.][]{vonderLinden2010}.

In this section we discuss the distribution of the galaxy populations within the cluster. The fraction of red, blue, star-forming and AGN as a function of the radial distance to the cluster centre provide clues about the relevance of the environment and/or AGN feedback in the quench of star formation. We also study the variation of the SFH parameters of galaxies that are in the central part ($r\leq 0.5$~R$_{200}$), with respect to outskirt regions ($r >0.5$~R$_{200}$); thus, the SFH-distance relation provides information about the accretion history and the differential quenching time scales. We finish the discussion with the variation of stellar population properties with cluster-centric radius; in particular,  the sSFR-distance relation traces how the environment-quenching process proceed.

\subsection{Spatial distribution of the galaxy populations}

The 2D map distribution of the galaxy populations of the cluster is shown in  Fig.~\ref{fig:spatial_distribution_with_contours}.  Most of the red galaxies are located inside the inner region (d$<$ $0.5~R_{200}$ from the BCG), while half of the blue galaxies are located around $0.9~R_{200}$. This visual inspection is corroborated by the mean distance of the RG, which is $0.60$~R$_{200}$, while for BG is around $0.98$~R$_{200}$, which is almost the same as the mean distance of ELGs ($0.90$~R$_{200}$, $0.98$~R$_{200}$ for the SF galaxies and $0.64$~R$_{200}$ for the AGNs). Moreover, the distribution of the ELGs is very similar to that of BG, because most of the ELGs are BGs. This indicates that RGs are more prominent in denser environments than BGs, as seen in previous works \citep[see e.g][]{Balogh2004}.  ELGs appear to show a more uniform distance distribution. This is in accordance with the results of the literature, such as \cite{Haines2012, Haines2015, Noble2013, Noble2016, Mercurio2021}. However, our ELG population is not composed exclusively of SF galaxies, there are also AGN candidates; this is justified by the presence of ELGs in denser environments. In fact, most of the AGNs are located in the central region, while the number of SF galaxies increases with distance.

\begin{figure*}
\centering
\includegraphics[width=\textwidth]{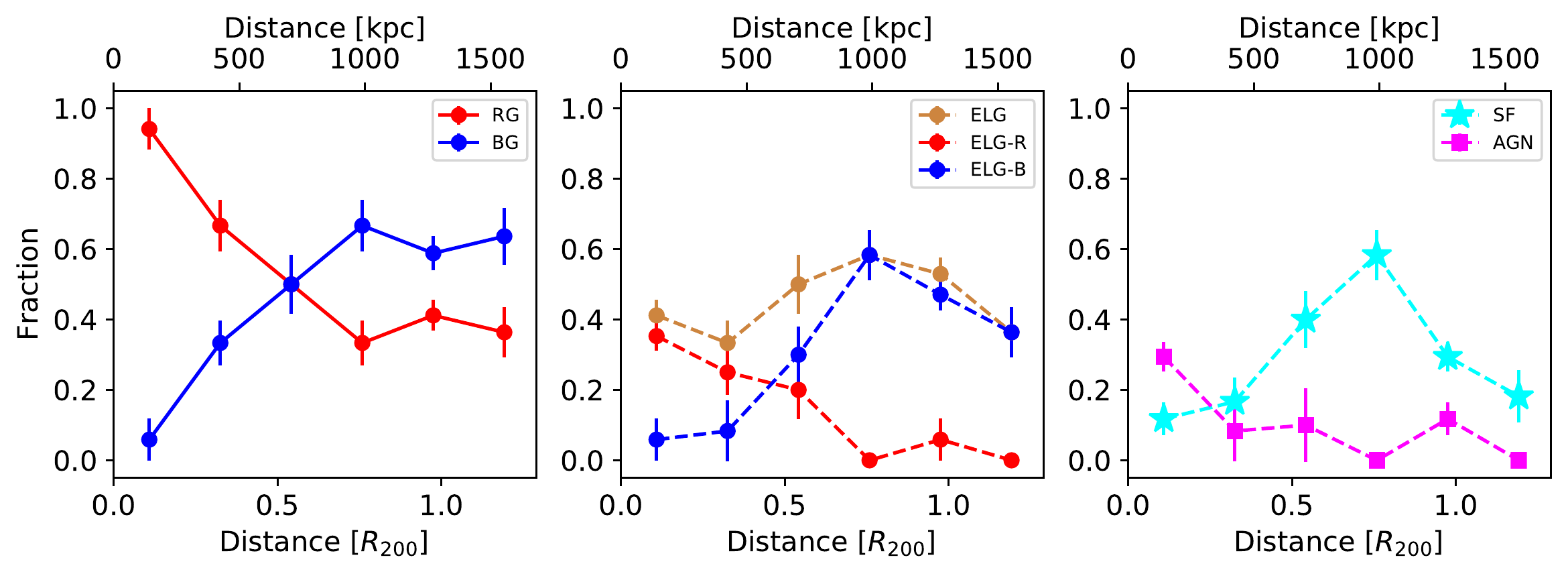}
\caption{First panel: radial variation of the fraction of red and blue galaxies. Red points represent the fraction of red galaxies, blue points represent the fraction of blue galaxies. Second panel: fraction of ELG, red ELG and blue ELG as a function of distance. Peru points represent the total fraction of ELG, red points the fraction of red ELG, and blue points the fraction of blue ELG. Third panel: fraction of SF galaxies and AGNs as a function of the distance to the BCG. Cyan stars represent the fraction of star forming galaxies. Magenta squares represent the fraction of AGN.}
\label{fig:fraction}
\end{figure*}

To quantify the spatial variation of the galaxy populations, we discuss how the fraction of the different galaxy populations changes with the distance to the BCG (see Fig.~\ref{fig:fraction}). Red galaxies clearly dominate over blue galaxies in the inner regions (up to $d\sim 0.5$~R$_{200}$). At this point, we are not affected by the incompleteness of the observations. The fraction of red galaxies steeply decreases, as the fraction of blue galaxies increases with the distance, but the fraction of red galaxies remains above the value obtained by \cite{Rosa2021} even at distances larger than R$_{200}$. It is also interesting to note that the fraction of red is equal to the fraction of blue galaxies at $d\approx0.5$~R$_{200}$. Thus, we can conclude that inside the virialised region, the red galaxy population dominated over the blue one. 

The fraction of ELG slightly increases with distance, up to $d\approx$~R$_{200}$ and then decreases again. This decrease of the ELG fraction for $d>$~R$_{200}$ could be a consequence of the possible observational incompleteness of our sample at larger distances. The fraction remains below $0.6$ at all distances. This fraction is  below the $\sim$80$\%$ star-forming field AEGIS population at $z=0.2-0.3$ \citep{Gines2022}.
If we separate blue and red ELG, we find that most of the ELG are red in the inner areas, but their fraction rapidly decreases and is negligible at distances larger than $d\approx0.5$~R$_{200}$. In contrast the fraction of blue ELG is negligible inside $d\approx0.3$~R$_{200}$, and then increases steeply. An almost equal behaviour is found for AGN and SF galaxy populations, respectively. As it happens for the blue ELG, the fraction of SF galaxies increases up to $0.6$ as the distance increases. Although it decreases to lower values at distances higher than $d\approx0.8$~R$_{200}$. This may be due to our incompleteness in the observations or to the presence of composite blue galaxies that cannot be clearly classified neither as SF nor as AGN.

Thus,  blue and star-forming galaxies are more common beyond the cluster virialised region, in agreement with other works \cite{Haines2012, Haines2015, Noble2013, Noble2016, Mercurio2021}. As well as \citet{Olave-Rojas2018}, we find that the fraction of red galaxies remains higher than in the field at distances larger than R$_{200}$. However, in contrast to \citet{Guglielmo2019}, we do not find a larger fraction of SF galaxies than blue galaxies, except for the most central region, where a reactivation of the SF may be taking place, due to the aforementioned mechanisms. The radial profile of the AGN fraction could compatible with the results by \citet{Peluso2022}, who find a significant abundance of AGNs in galaxies that have suffered ram-pressure stripping, taking into account the relation between ram-pressure stripping and the ICM density AGN feedback might play a role at the centre due to the increase of AGN fraction toward it; but it could not explain the large fraction ($>$ 50$\%$) of red galaxies inside 0.5~$R_{200}$. Other environment-related processes plus possible previous mass-quenching may be acting in the cluster.

\subsection{Stellar population properties: radial variation of the colours, mass and ages}

\begin{figure*}
\centering
\includegraphics[width=\textwidth]{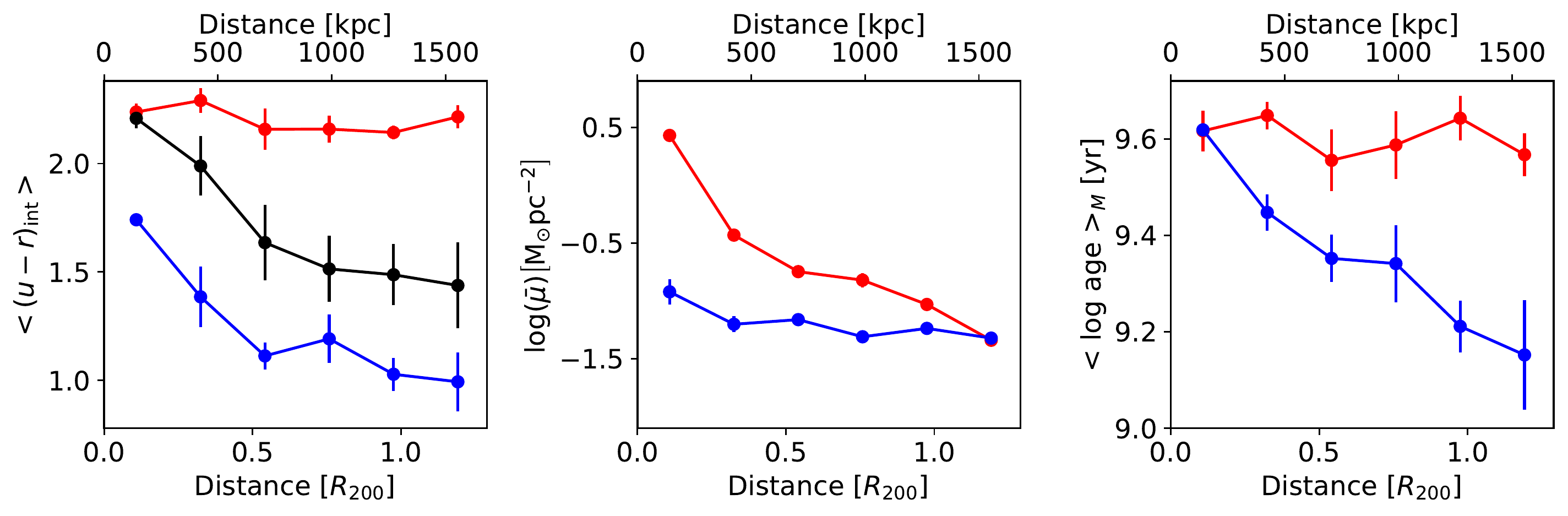}
\caption{Radial distribution of the stellar mass surface density, the mean mass-weighted age and the mean $(u-r)_\mathrm{int}$ colour. The grey dashed lines represent the limit of the FoV of \mjp. Red dots represent the values for the red galaxies and blue dots represent the values for the blue galaxies. Black dots represent the values of $(u-r)_\mathrm{int}$ for all the galaxies.
}
\label{fig:mass_age_radial}
\end{figure*}

The variation of the abundance of red and blue galaxies has an important effect in the radial variation of the galaxy properties, in particular in the colours, and ages of the stellar populations. 
 
We study the radial profiles of the $(u-r)_\mathrm{int}$ colour, the mass density (with the area corrected from mask an incompleteness effects) and the <log age>$_M$ (see Fig.~\ref{fig:mass_age_radial}). The colour, $(u-r)_\mathrm{int}$, decreases with the radial distance; however, this is in part due to the radial variation of the fraction of red and blue galaxies in the cluster. When the galaxy population is segregated in red and blue galaxies, we find that most of the red galaxies have very similar $(u-r)_{int}$, while blue galaxies are bluer from inside-out of the cluster centre. This is probably associated to a change in the age of the blue galaxy population. Red galaxies ages stay approximately constant at <log age>$_M\approx 9.6$. \ \footnote{This is the logarithm of the stellar ages expressed in yr}. Meanwhile, the mean age of BGs decreases as the distance to the BCG increases by about $0.3$~dex in the inner $0.5$~R$_{200}$. The stellar mass surface density of the red galaxies decreases with the distance to the BCG, reflecting that the most massive galaxies are sited in the inner $0.5$~R$_{200}$. Blue galaxies show a lower mass density and a smoother slope than RGs, but they show similar stellar mass density to the red galaxies beyond $0.5$~R$_{200}$.

Therefore, we find that red, more massive, older galaxies, are found in the inner areas while we find, blue, younger, and less massive galaxies in the outskirts. These properties are generally associated to galaxies in the red sequence (which also, show low values of the star formation) and the blue cloud with usually higher star formation than most galaxies in the red sequence\citep[see e.g.][]{Kauffmann2003a, Kauffmann2003b, Baldry2004, Brinchmann2004, Gallazzi2005, Mateus2006, Mateus2007}.

The stellar ages of the blue galaxies show a clear gradient with cluster-centric distance. However, the mean ages of the red galaxies is almost constant with the radial distance. It suggests that these galaxies were probably quenched earlier than their accretion to the cluster or during the first epoch of the accretion.

\subsection{SFH: Spatial variation}
\label{sec:SFH}

\begin{table*}
    \centering
    \caption{ The average and dispersion for the SFH parameters of the red and blue galaxy members, and for galaxies inside $0.5$~R$_{200}$, between $0.5$~R$_{200}$ and R$_{200}$ and outside R$_{200}$. }
    \begin{tabular}{c c c c c c c c c}
    \hline
    \hline
        {\small SP} & {\small RG} &  {\small BG}  &  {\small RG ($d<0.5$)}& {\small BG ($d<0.5$)} & {\small RG ($0.5<d<1$)} &  {\small BG ($0.5<d<1$)} & {\small RG ($d>1$) } &   {\small BG ($d>1$)} \\
     \hline
$t_0$ & $6.44 \pm 1.76$ & $6.00 \pm 1.72$ & $6.64 \pm 1.75$ & $7.69 \pm 0.17$ & $6.30 \pm 1.86$ & $6.37 \pm 1.59$ & $6.18 \pm 1.65$ & $5.33 \pm 1.66$\\
$\tau/t_0$ & $0.12 \pm 0.02$ & $0.94 \pm 0.63$ & $0.12 \pm 0.03$ & $0.46 \pm 0.21$ & $0.12 \pm 0.02$ & $0.83 \pm 0.49$ & $0.11 \pm 0.02$ & $1.14 \pm 0.70$\\
        \hline
    \end{tabular}
    
    \label{tab:SFHparam}
\end{table*}

\begin{figure*}
\centering
\includegraphics[width=\textwidth]{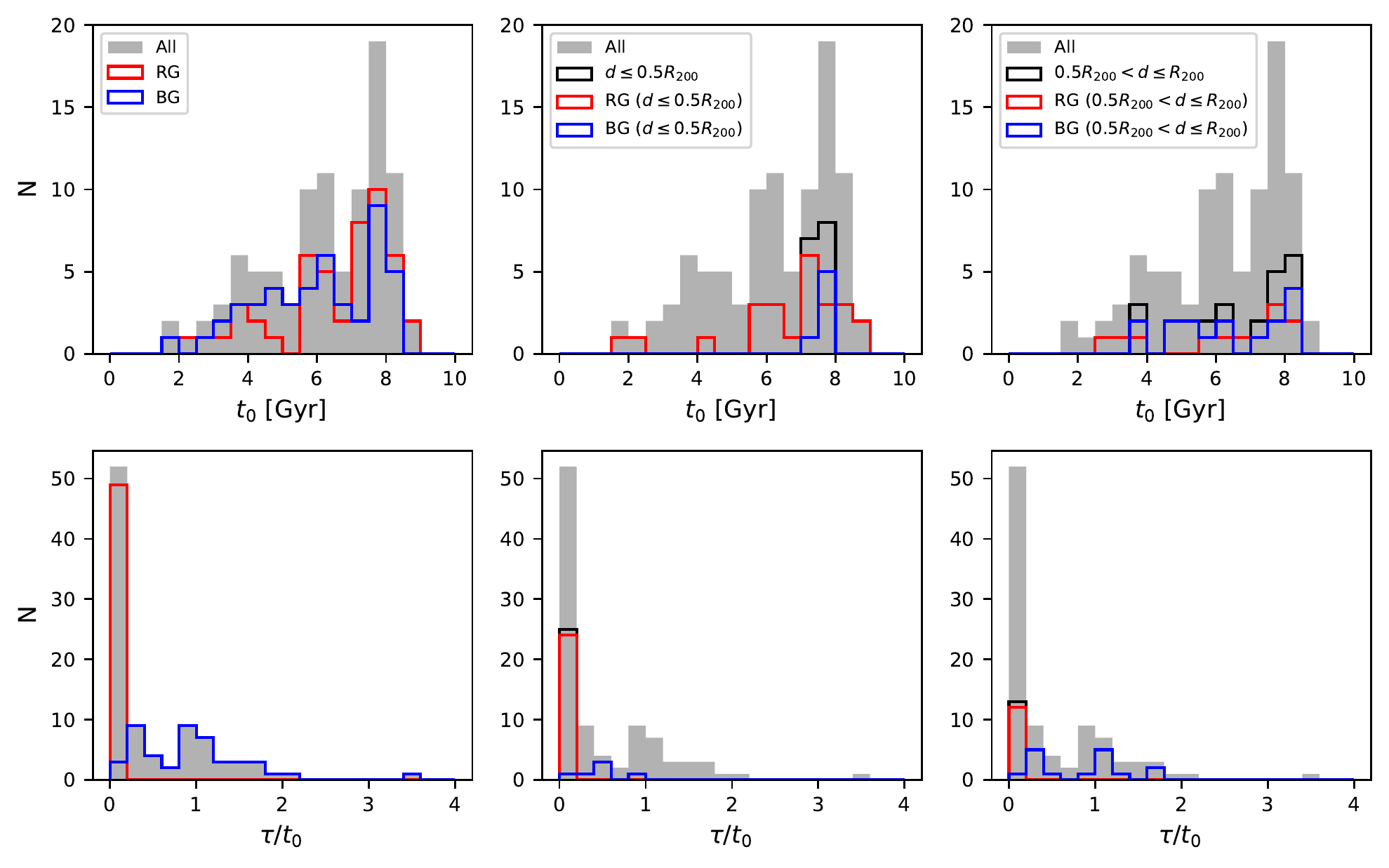}
\caption{From left to right: Distributions of SFH parameters t0 (top row) and  $\tau$/t0 (bottom row). Grey histograms represent the distribution of all the galaxies in the cluster in all panels. Red, blue and black histograms represent the distribution of red, blue and all galaxies at different cluster-centric distances: All the galaxies in the cluster (first column), galaxies within $0.5$~R$_{200}$ (Second column) and galaxies within $0.5$~R$_{200}$ and R$_{200}$ (third column)
}
\label{fig:hist_t0_tau}
\end{figure*}

We now investigate the spatial variation of the SFH of the cluster galaxies. Before doing that, however, let us comment on the uncertainties involved.

The reliability of our methodology has been previously assessed in \citet{Rosa2021}, where we have shown that the SFH of a complete sub-sample of \mjp{} galaxies selected at $z \sim 0.1$ constrains the cosmic evolution of the star formation rate density up to $z \sim 3$, producing results in good agreement with those derived from cosmological surveys. We have further shown that the galaxy properties (stellar mass, ages and metallicity) inferred by fitting the \js{} with the non-parametric codes \muff{}, \alstar{}, and \tgas{}, are similar to those obtained by \baysea{} using the same delayed-$\tau$ model used in this work (see Table 1 in \citealt{Rosa2021}).

Reassuring as these statistical results are, one should not loose sight of the inherent difficulties in estimating SFHs of individual galaxies \citep[e.g.]{ocvirk2006}. Our Bayesian analysis based on an analytical prescription for the SFH (eq.\ \ref{eq:SFH}) is just one out of a vast spectrum of alternative approaches. Parametric models such as our delayed-$\tau$ model are known to lack flexibility to emulate the diversity of SFHs in galaxies \citep[e.g.]{dressler2016,pacifici2016} and to induce considerable biases in some cases \citep[e.g]{lower2020}. Non-parametric models alleviate these problems, but the higher dimensionality associated with the added flexibility requires extra care when specifying the priors, which may have a significant impact in the estimated properties \citep{leja2019}. Moreover, it has been previously shown that only a few characteristic episodes in the SFH can be retrieved from the SED-fitting \citep{cid2005, ocvirk2006, asari2007,  tojeiro2017}. 

Despite all caveats involved, our previous analysis of $\sim$8000 \mjp{} galaxies, where we detected only small differences between properties derived though parametric and non-parametric codes (significantly below the 0.4 dex in \logM{} inferred for the SED-fitting of broad band photometry of mock data of cosmological galaxy formation simulations by \citealt{lower2020}), gives us confidence that we can use our results to investigate the general trend of the spatial variation of the SFH among the galaxy cluster members. The comparative and statistical nature of this analysis further alleviates worries associated with the SFH parameters derived for each galaxy, which, as discussed above, should be treated with caution.

We focus on two parameters for this study: $t_0$, the lookback time when the star formation began, and $\tau/$t$_0$, a measure of the extent of the star formation that is better constrained than $t_0$ or $\tau$. Table \ref{tab:SFHparam} and Fig.\ \ref{fig:hist_t0_tau} summarise our results. We divide the galaxies into blue and red ones once again. We further divide galaxies by their (projected) distances to the BCG into smaller than $0.5$~R$_{200}$, between $0.5$~R$_{200}$ and $1$~R$_{200}$, and larger than R$_{200}$ bins. This allows us to distinguish the effect of the environment in the SFH for the central virialised area and the outer regions.

The parameter $t_0$ shows similar values for red and blue galaxies at all  distances. This would suggest that most galaxies started forming stars roughly at the same epoch (around $\sim 6.5$~Gyr). The main differences appear in the blue galaxies within and outside of 0.5$~R_{200}$. Blue galaxies in the inner region show the highest mean value of $t_0$, but they are very few and this value is compatible with the one obtained for red galaxies at similar distances. Blue galaxies in the outer region ($d >$~R$_{200}$) show a lower value of $t_0$. This could be a consequence of these galaxies being in the cluster infall region \citep{Rines2006}.

Since values of $t_0$ are very similar for most galaxies all over the cluster, we interpret the low values of $\tau/\mathrm{t}_0$ as short episodes of star formation and large values of this parameter as star formation more extended in time. Red galaxies values of $\tau/\mathrm{t}_0 \sim 0.12$ no matter their distance to the BCG,  suggesting that their star formation was shut down very fast. On the contrary, blue galaxies show larger values of this fraction than red galaxies at all distances, and there is a clear increase in $\tau/\mathrm{t}_0$ as the distance to the BCG increases.

Thus, these results suggest a faster quenching process for blue galaxies in the dense (inner) regions;  while red galaxies might be quenched earlier on, and independently of the distance to the cluster centre in an earlier cluster accretion epoch. Moreover, the quenching of the star formation of red galaxies might be linked to the AGN and/or galaxy stellar mass rather than to the environment because at the smaller cluster-centric distance is where we find the most massive galaxies and the fraction of AGN is larger.

\subsection{sSFR: radial variation}

\begin{figure}
\centering
\includegraphics[width=0.45\textwidth]{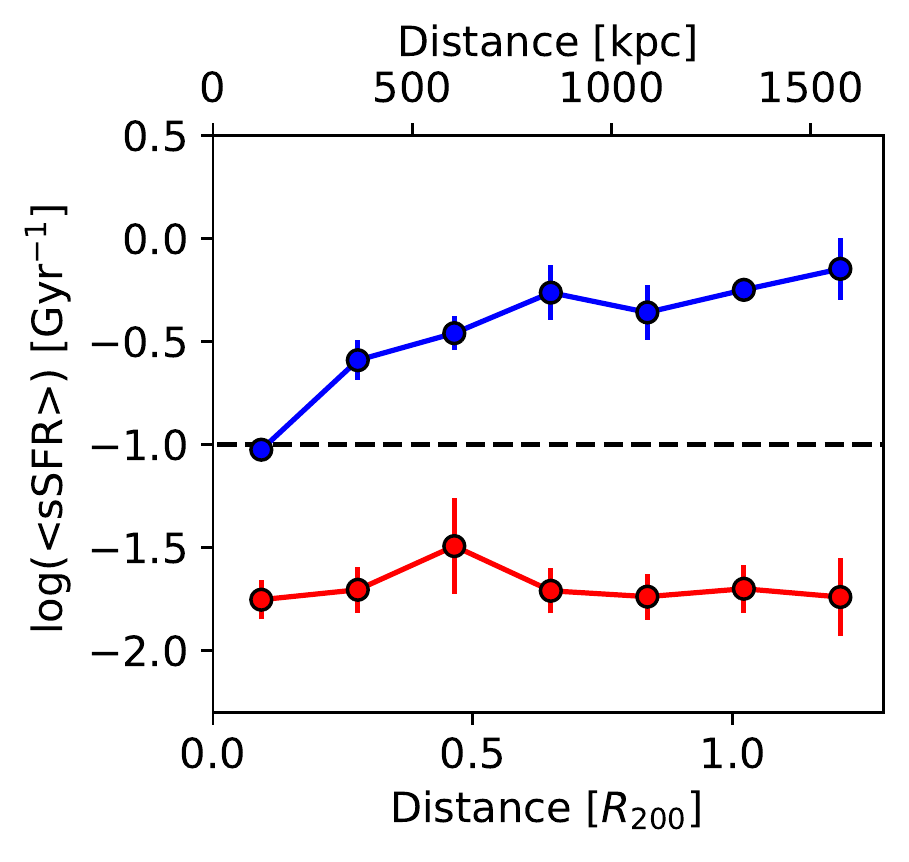}
\caption{Radial distribution of the mean sSFR. The gray dashed lines shows the limit of the FoV of \mjp. The black dashed line shows \cite{Peng2010} criteria to distinguish among star forming and quiescent galaxies. Red dots represent red galaxies. Blue dots represent  blue galaxies. }
\label{fig:sSFR}
\end{figure}

From the SFH parameters we can conclude that red galaxies have already quenched independently of their position in the cluster; while for blue galaxies the quench process proceeds from the inner to outer regions of the cluster. To confirm this conclusion, we study the variation of the sSFR with the distance (see Fig.~\ref{fig:sSFR}). It is worthy to notice that blue galaxies have a sSFR at all the cluster-centric radius that are above $sSFR=0.1$~Gyr$^{-1}$, which is the threshold adopted by \cite{Peng2010} to differentiate star forming galaxies from quenched galaxies. The mean sSFR of blue galaxies also clearly increases from inside-out of the cluster; in contrast, the mean sSFR of the red galaxies remains constant within the error bars. This  suggests  that red galaxies have quenched before their accretion to the cluster or were quenched within it in an earlier accretion epoch, while blue galaxies are still in the process of quenching. This could be related to the pre-processing effects during the infalling processes \citep[assuming galaxies are incorporated in substructures already evolving, see e.g.][]{Gavazzi2003, Aguerri2017, Donnari2021}. Our results are differ from those of \citet{Knowles2022}, who find no dependence of the star formation with the cluster-centric distance, but the distinction between red/blue galaxies is key for this result. Also, in particular, \cite{Balogh1999} results (in a redshift range similar to our cluster)  show that the last episode of star formation is more recent for galaxies in the outskirts than in inner regions.

\subsection{On the pre-processing scenario}

To sum up, all the galaxies were formed around the same epoch (with the exception of some outer blue ones). Red galaxies had shorter star formation periods and have a similar SFH, independently of their position in the cluster. Meanwhile, blue galaxies are still forming stars or have been forming them until very recently, and galaxies in the inner regions are quenching faster than in the outer ones. However, red galaxies were quenched earlier on, independently of their position on the cluster. These  results suggest different evolutionary paths and accretion histories for red and blue galaxies.

Illustris cosmological simulations have shown that pre-processing plays a relevant role in quenching galaxies \citep{Donnari2021}. They find that satellites can be quenched before infalling in dense environment, or after being accreted into any host; or while being members of pre-processing hosts other than the actual one where they are found today. AGN feedback and mass-quenching may be acting in the pre-processing host phases. This is a possible scenario for explaining the spatial variation of the SFH of red and blue galaxies, their abundance, AGN fraction, and variation of the galaxies properties with the radial distance to the cluster centre.

Another scenario is the 'slow-then-rapid' quenching \citep[see e.g.][]{Maier2019, Roberts2019, Kipper2021}, where galaxies suffer slow quenching processes and once they enter a dense environment, start a faster quenching phase. Results from \cite{Pallero2022}, using the \texttt{C-EAGLE} simulation support this scenario. They show that these processes usually become relevant at $\sim \mathrm{R}_{200}$, where the ICM reaches a density high enough for ram-pressure stripping to  become relevant. They also find that the fraction of galaxies quenched  {\it in situ} in comparison the fraction of galaxies quenched because of pre-processing decreases as $M_{200}$ increases. 

A combination of both scenarios may serve us to interpret the spatial and radial distributions of the stellar population properties that we find. We find red galaxies with very similar properties along the whole cluster. Red galaxies may quenched inside smaller structures, that were later accreted to the cluster.  According to \citet{Pallero2022} results, for a cluster of this mass ($M_{200}=3.3\times 10^{14} ~ \mathrm{M}_\odot$), we should expect a similar fraction of galaxies quenched inside the cluster and quenched via pre-processing. However, we find that $\sim 73$~\% of the red galaxies are inside R$_{200}$, so in order to be compatible with these results some of the inner red galaxies must be part of a different halo. On the other hand, \citet{Donnari2021} results show that the pre-processing scenario is relevant for low mass galaxies, and that massive galaxies quench on their own.  

If we assume that some of the blue galaxies belong to the original halo and some have been accreted later, we could explain the behaviour of blue galaxies and the larger dispersion of their properties, and the larger quenching of inner ones. However, \citet{Pallero2022} estimate that the quenching timescale for galaxies once the in-fall beyong R$_{200}$ is $\sim 1$~Gyr, but our estimations of $\Delta t_q$ are larger for blue galaxies, and only some of the red galaxies are compatible with these value, regardless of their distance to the cluster centre. These suggest that the accretion and evolution scenario may be more complex and a different model is required.

\section{Summary and conclusions}
\label{sec:conclus}
In this paper we have studied the stellar population properties of the \mjp \ cluster mJPC2470-1771 using the \jp \ photometric filter system. Its redshift is $z=0.29$, its mass $M_{200}=3.3\times 10^{14} ~ \mathrm{M}_\odot$ and its radius $R_{200}=1.304 ~ \mathrm{kpc}$.  We used the fossil record method for stellar populations, we analyse the SEDs (\js) of the galaxy members of the cluster. It was detected and the members were selected using AMICO (Maturi et al., in prep.) using the \photozbest{}, selecting a total of 99 objects.

We have used the \baysea \ code to fit the stellar continuum and constrain the stellar population properties, by assuming a delayed-$\tau$ model for the SFH. 
The parameters obtained with \baysea \ are the stellar  mass,  the  metallicity,   the  extinction $A_V$, $t_0$,  and $\tau$. We used these parameters and fittings to calculate the mass- and light-weighted ages and the extinction-corrected rest frame $(u-r)$ colours. 
We have established a criteria to select the ELG population, using the median error of the closest filters to H$\alpha$ wavelength, and the predictions of the EW(H$\alpha$), EW([NII]), EW(H$\beta$) and EW([OIII]) made with \cite{Gines2021} ANN. We use the WHAN and BPT diagrams to separate SF, AGNs and quiescent-retired galaxies.
We studied the spatial distribution of the stellar population properties in the cluster, and the radial distribution of the abundances of red, blue, SF galaxies and AGN hosts.

The main conclusions after all our analysis are:

\begin{itemize}
    \item{We observe a fraction of red galaxies (52~\%) that is larger than that in the whole AEGIS field set of galaxies with redshift $0.25<z<0.35$, which is $\sim 20\%$. The distribution of the stellar population properties in the mass-colour diagrams is the same as the whole set.}
    
    \item{We have selected a total of 48 ELG. They are dominated by young galaxies and most of the blue, less massive galaxies are selected as ELG. There are red galaxies in this set, showing the lowest inferred values of EW(H$\alpha$), being the median value equal to 8.96 \AA. 65.3\% of these galaxies are probably star forming galaxies, 24.4\% could be AGNs and the rest could be SF, AGNs or composite galaxies.}
    
    \item{The red, older, more massive galaxies are mainly located in the inner part ($d< 0.5$~R$_{200}$) of the cluster, where the density is higher. The blue, and SF galaxies are more numerous at ($d> 0.5$~R$_{200}$), and their abundance increases with radial distance, being equal to the red galaxy fraction at $d\sim 0.5$~R$_{200}$. The abundance of the AGNs population decreases with the radial distance, and it is higher at the cluster centre.}
    
    \item{Analysing the SFH, we find that galaxy members were formed roughly at the same epoch, but blue galaxies have had more recent star formation periods. Our results are compatible with a scenario where red galaxies are quenched prior to the cluster accretion or an earlier cluster accretion epoch. While blue galaxies may be in the transition to be quenched. This is also supported by the radial distribution of the red and blue galaxy populations, because the mean stellar age remains constant for red galaxies, but decreases for blue galaxies with the distance to the BCG.} 
       
    \item{The sSFR of the red galaxies is almost constant with radial distance at sSFR $\sim 0.02$ Gyr$^{-1}$. The sSFR of blue galaxies decreases with the cluster-centric radius from sSFR $\sim 0.1$~Gyr$^{-1}$ to above $\sim 0.7$~Gyr$^{-1}$ beyond $0.5$~R$_{200}$.  This suggests that the quenching of blue galaxies is progressing from the inside-out of the cluster.
    AGN feedback and/or mass might also be intervening in the quenching of red galaxies.}

\end{itemize}


\begin{acknowledgements} 

R.G.D., L.A.D.G., R.G.B., G.M.S., J.R.M., E.P. acknowledge financial support from the State Agency for Research of the Spanish MCIU through the "Center of Excellence Severo Ochoa" award to the Instituto de Astrof\'\i sica de Andaluc\'\i a (SEV-2017-0709), and to the AYA2016-77846-P and PID2019-109067-GB100.

G.B. acknowledges financial support from the National Autonomous University of M\'exico (UNAM) through grant DGAPA/PAPIIT IG100319 and from CONACyT through grant CB2015-252364.

SB  acknowledges PGC2018-097585- B-C22, MINECO/FEDER, UE of the Spanish Ministerio de Economia, Industria y Competitividad. 

L.S.J. acknowledges support from Brazilian agencies FAPESP (2019/10923-5) and CNPq (304819/201794).

P.O.B. acknowledges support from  the Coordenação de Aperfeiçoamento de Pessoal de Nível Superior – Brasil (CAPES) – Finance Code 001.

P.R.T.C. acknowledges financial support from Funda\c{c}\~{a}o de Amparo \`{a} Pesquisa do Estado de S\~{a}o Paulo (FAPESP) process number 2018/05392-8 and Conselho Nacional de Desenvolvimento Cient\'ifico e Tecnol\'ogico (CNPq) process number  310041/2018-0. 

V.M. thanks CNPq (Brazil) for partial financial support. This project has received funding from the European Union’s Horizon 2020 research and innovation programme under the Marie Skłodowska-Curie grant agreement No 888258.

E.T. acknowledges support by ETAg grant PRG1006 and by EU through the ERDF CoE grant TK133.

Based on observations made with the JST/T250 telescope and PathFinder camera for the miniJPAS project at the Observatorio Astrof\'{\i}sico de Javalambre (OAJ), in Teruel, owned, managed, and operated by the Centro de Estudios de F\'{\i}sica del  Cosmos de Arag\'on (CEFCA). We acknowledge the OAJ Data Processing and Archiving Unit (UPAD) for reducing and calibrating the OAJ data used in this work.

Funding for OAJ, UPAD, and CEFCA has been provided by the Governments of Spain and Arag\'on through the Fondo de Inversiones de Teruel; the Arag\'on Government through the Research Groups E96, E103, and E16\_17R; the Spanish Ministry of Science, Innovation and Universities (MCIU/AEI/FEDER, UE) with grant PGC2018-097585-B-C21; the Spanish Ministry of Economy and Competitiveness (MINECO/FEDER, UE) under AYA2015-66211-C2-1-P, AYA2015-66211-C2-2, AYA2012-30789, and ICTS-2009-14; and European FEDER funding (FCDD10-4E-867, FCDD13-4E-2685).

Partially based on observations obtained at the international Gemini Observatory, a program of NSF’s NOIRLab, which is managed by the Association of Universities for Research in Astronomy (AURA) under a cooperative agreement with the National Science Foundation. on behalf of the Gemini Observatory partnership: the National Science Foundation (United States), National Research Council (Canada), Agencia Nacional de Investigaci\'{o}n y Desarrollo (Chile), Ministerio de Ciencia, Tecnolog\'{i}a e Innovaci\'{o}n (Argentina), Minist\'{e}rio da Ci\^{e}ncia, Tecnologia, Inova\c{c}\~{o}es e Comunica\c{c}\~{o}es (Brazil), and Korea Astronomy and Space Science Institute (Republic of Korea).

R.A.D. acknowledges support from the Conselho Nacional de Desenvolvimento Científico e Tecnológico - CNPq through BP grant 308105/2018-4.

Funding for the J-PAS Project has been provided by the Governments of Spain and Aragón through he Fondo de Inversión de Teruel, European FEDER funding and the Spanish Ministry of Science, Innovation and Universities, and by the Brazilian agencies FINEP, FAPESP, FAPERJ and by the National Observatory of Brazil. Additional funding was also provided by the Tartu Observatory and by the J-PAS Chinese Astronomical Consortium.

We acknowledge the technical members of the UPAD for their invaluable work: Juan Castillo, Tamara Civera, Javier Hernández, Ángel López, Alberto Moreno, and David Muniesa.

We thank Rain Kipper, Roberto Soria, Juan Antonio Fern\'{a}ndez Ontiveros and Ana M. Conrado for their comments and suggestions to improve the final paper.

We deeply thank the referee for their comments and suggestions that have greatly improved the paper.

This paper has gone through internal review by the J-PAS collaboration.

\end{acknowledgements}




\bibliographystyle{aa}
\bibliography{bibliography.bib}


\begin{appendix}

\section{AMICO versions}
\label{appendix:AMICOversions}

\begin{figure}
    \centering
    \includegraphics[width=0.5\textwidth]{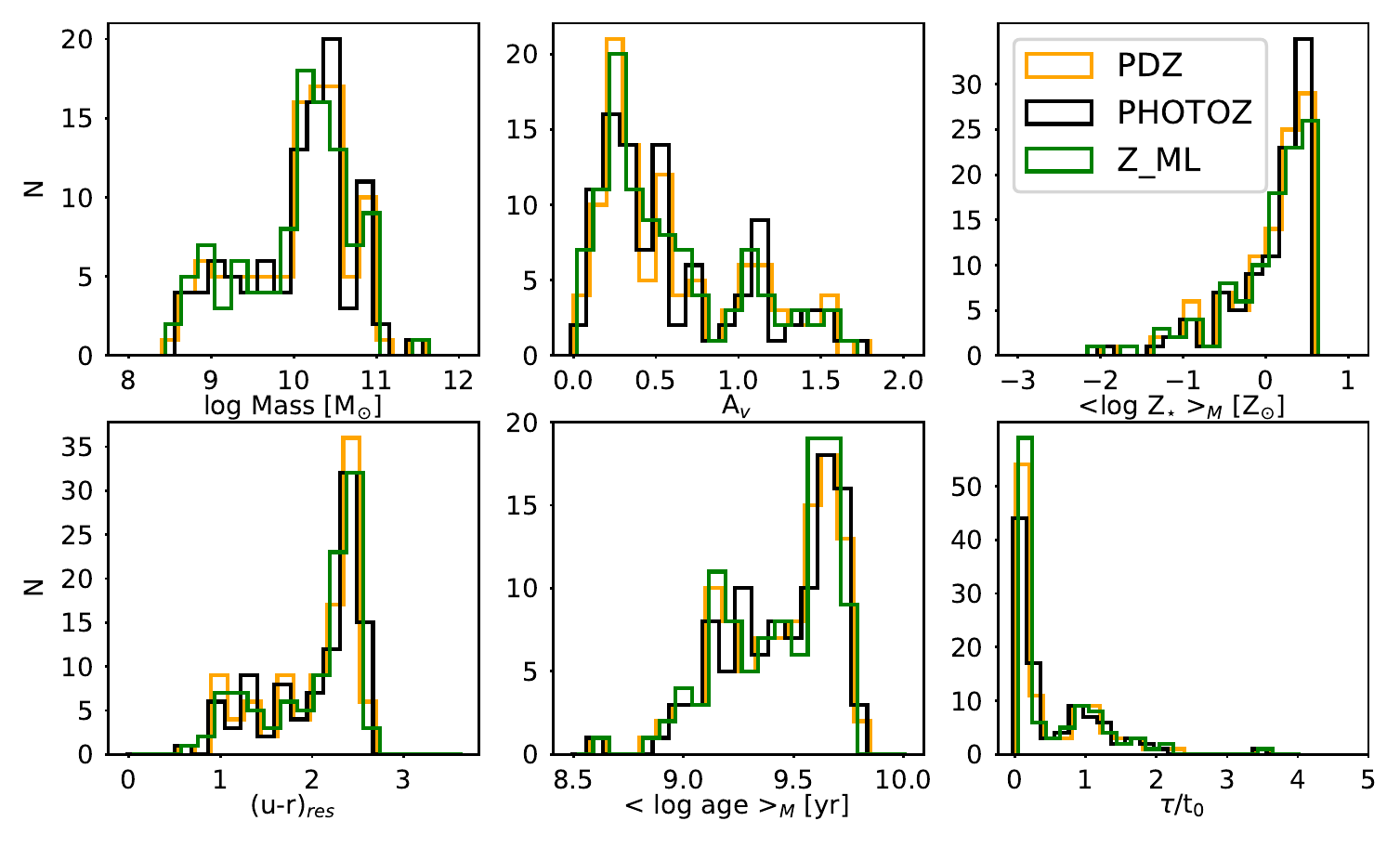}
    \caption{Stellar population properties of the galaxies belonging to the cluster, using different AMICO catalogues.}
    \label{fig:AMICO_comparison_05}
\end{figure}

As mentioned in Sect.\ref{sec:galaxymembers}, at the moment of writing of this paper there are three versions of the code AMICO in study. The difference resides in the redshift used for the computation of the amplitude, the association probability and the rest of the parameters. One uses the best \photozbest{} (the absolute maximum of the probability density function of the redshift or $z\mathrm{PDF}$) as input, other version uses $z_{\mathrm{ml}}$ (the median redshift of the $z\mathrm{PDF}$) the last one uses the $z\mathrm{PDF}$ itself (see \citealt{HC2021} for more information about the redshifts).  The final version  (including a group catalog for \mjp) will be published in a future paper (Maturi et al., in preparation). Since those results and the details about AMICO escape the scope of this work, we refer the reader to \cite{2005A&A...436...37M, AMICO}, and restrict ourselves to the study of the stellar population properties.

The parameter  we use to select our set of galaxies is the association probability. When changing the redshift, so does the association probability. In consequence, some galaxies may be considered members or not of the cluster. In fact, the total number of galaxies changes (99 galxies using the \photozbest{}, 95 using the $z_{\mathrm{ml}}$ and 114 using the $z\mathrm{PDF}$. Not only the number of galaxies changes, but also the set of the common galaxies between two different versions of the code does not equal the smallest set of the two, although they are very similar overall. There are 84 galaxies in common among the three sets, and the larger discrepancies are  12 galaxies that appear with \photozbest{} and $z\mathrm{PDF}$ but not with $z_\mathrm{ml}$, and other 12 galaxies that only appear with $z\mathrm{PDF}$ .

In Figs. \ref{fig:AMICO_comparison_05} we show the distribution of the stellar properties obtained with the three different versions. These distributions are practically the same, and the differences that appear are mainly due to the different number of galaxies in each set, but the whole image of the cluster remains the same. This is an important result for us, since our results remain valid independently of the final version that may be chosen, and for AMICO, since it proves that the cluster catalogue is robust regardless of the redshift definition being used.

\section{Comparison of \jp \ data with GMOS spectroscopy.}
\label{appendix:GMOS}

\begin{figure*}%
    \centering
    \subfloat{{\includegraphics[width=0.45\textwidth]{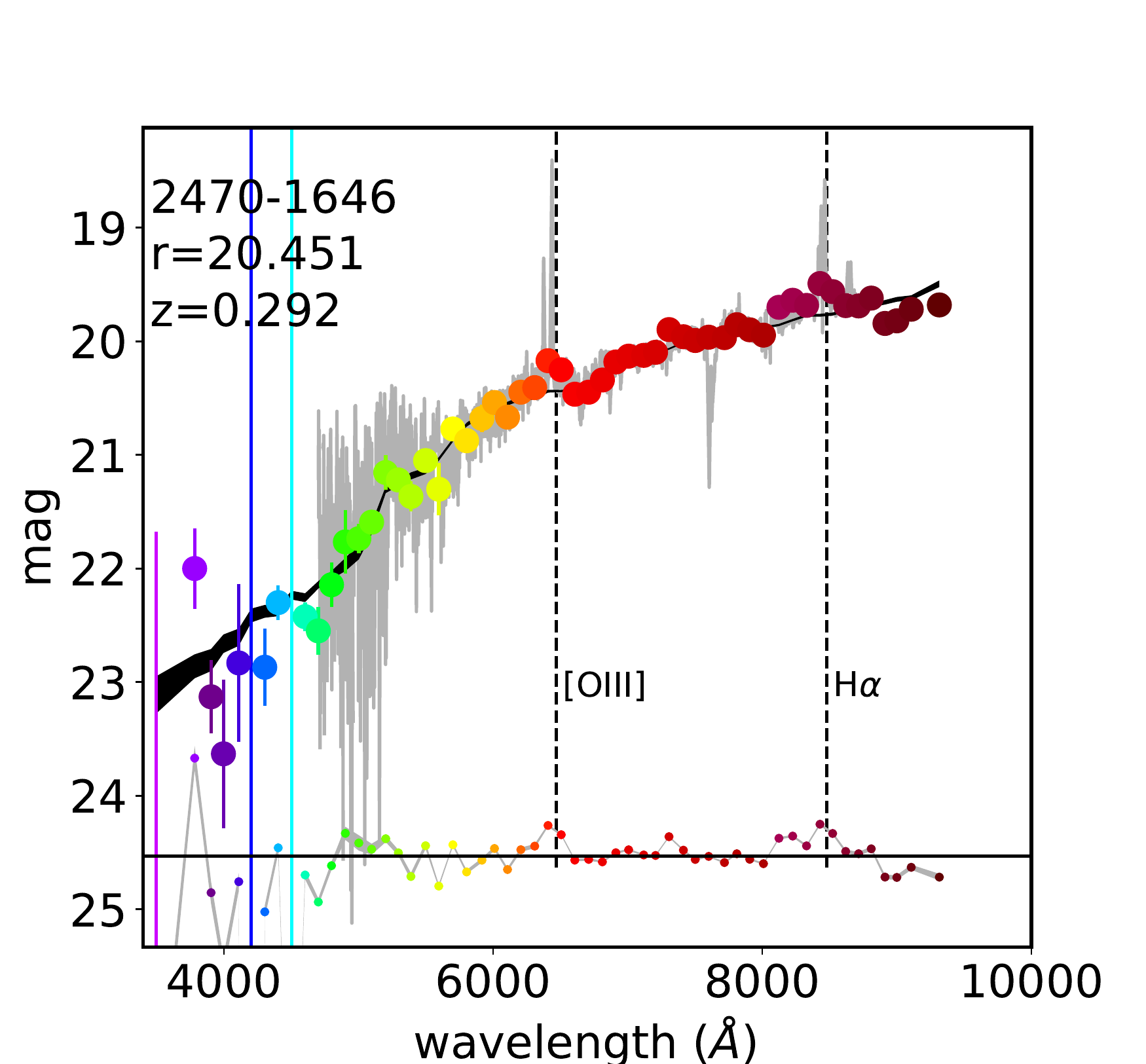} }}%
    \qquad
    \subfloat{{\includegraphics[width=0.45\textwidth]{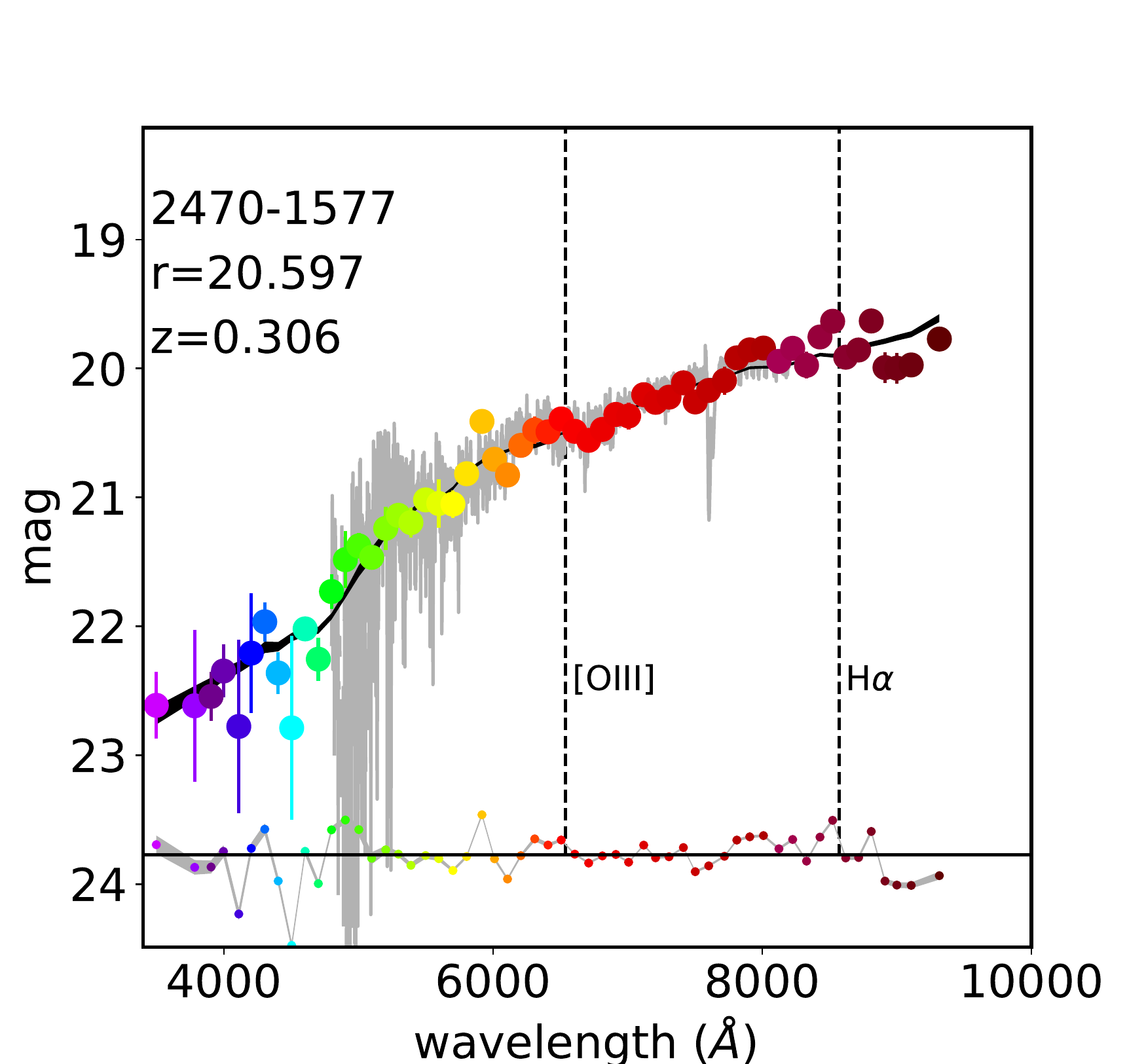} }}%
    \caption{Comparison among the \js \ and the GMOS spectra of two galaxies belonging to the cluster. Colour points represent the \js{}. Grey lines represent the GMOS spectra. The black band shows the magnitudes of the mean model $\pm$ one $\sigma$ uncertainty level. The difference between the model and the mean model fitted magnitudes are plotted as a small coloured points around the black bottom line. Vertical, black dashed lines show the wavelengths corresponding to [OIII] and H$\alpha$ emission lines.}%
    \label{fig:GMOS_JPAS}%
\end{figure*}

\begin{table*}[]
    \centering
    \caption{Emission line classification methods results. First column shows the \jp \ ID of the galaxies with H$\alpha$ emission in GMOS. Second and third columns show if the galaxy was classified as an emission line one with the median error method and the ANN method, respectively. Fourth and fifth column show some comments on why each method failed to classify the galaxy properly.}
    
    \begin{tabular}{p{0.15\textwidth} p{0.1\textwidth} p{0.1\textwidth} p{0.25\textwidth} p{0.25\textwidth}}
         \hline
         \hline
         Galaxies with H$\alpha$ emission in GMOS & Median error classification & ANN classification & Median error notes&  ANN notes\\
         \hline
         2470-1168 & Yes & Yes& - & -\\ 
         2470-494 & Yes & Yes& - & -\\ 
         2470-536 & ? & ?& J-spectra shows no emission & Unable to calculate values\\ 
         2470-1646 & Yes & Yes& - & - \\
         2470-1744 & No & No& Classified as line-emission when using the spectroscopic redshift & EW$_{\mathrm{min}}$=35.31, EW=2.44$\pm$6.32\\ 
         2470-2129 & Yes & Yes& - & -\\ 
         2470-2328 & Yes & Yes& - & -\\ 
         2470-2524 & Yes & Yes& - & -\\ 
         2470-2401 & Yes & Yes& - & -\\
         2470-2667 & ? & ?& J-spectra shows no emission & Unable to calculate values\\ 
         2470-1920 & Yes & Yes& - & -\\ 
         2470-1625 & Yes & Yes& - & -\\ 
         2470-2403 & Yes & Yes& - & -\\ 
         \hline
    \end{tabular}
    \label{tab:GMOS_ha_emission_summary}
\end{table*}

We did a spectroscopic follow up of the cluster with the Gemini Multi Object Spectrometers \citep[GMOS][]{Hook2004} mounted on the Gemini North telescope (Gemini program ID: GN-2020A-DD-203, PI: Carrasco). In total, we measured the spectroscopic redshifts for 53 galaxies observed with GMOS, of which 38 galaxies are members of the cluster. Figure~\ref{fig:GMOS_JPAS} shows a comparison among the \js{} and the spectra obtained from GMOS for two galaxies belonging to the cluster. This comparison shows the power of \jp{} photometric system to provide information equivalent to spectroscopy data.
We have already shown that \js{} can retrieve the stellar population properties with similar precision to spectroscopic datasets with S/N$\geq$10 \citep{Rosa2021}. With respect to GMOS data, \js{} have a better S/N ratio and it covers a larger wavelength range that is not affected by calibration issues at the wavelengths limits. Furthermore, many of the GMOS spectra do not cover H$\alpha$ wavelength range; and in only 7 galaxies of them H$\alpha$ is observed. The right panel of  Fig.~\ref{fig:GMOS_JPAS} shows an example of a galaxy with measured H$\alpha$ emission through the ANN, and clearly detected in the \js, that is not covered in the GMOS spectrum. In addition, \mjp{} data allowed us to observe most of the galaxies of the cluster brighter than 22.5 magnitude in the \rb{} band. In contrast, MOS spectroscopy is limited by the minimum distance between fibers, that prevent the simultaneous observations of galaxies that are close in the sky, and some fibers are contaminated by several close objects. Thus, \jp \ data is more suitable than GMOS spectra for our analysis. 

In addition to the galaxies in the clusters, our GMOS observations include more galaxies of \mjp{}. To test our ELG detection methods we include all the galaxies from \mjp{} that were observed in the $H\alpha$ wavelength range. This includes 7 spectra of the galaxies belonging to the cluster plus another 6 galaxies outside the cluster that show H$\alpha$ emission in their GMOS spectra.  Table~\ref{tab:GMOS_ha_emission_summary} summarises the results obtained with both methods for these galaxies. We note that none of the methods classify as ELG any of the galaxies that show no H$\alpha$ emission in their spectra (covering the corresponding wavelength). The different criteria for red and blue galaxies is proven to be required in this table. If we choose the same $\sigma=1$ detection level for blue or red galaxies, we find that red galaxies that actually show no emission line are classified as ELG.  If we choose the same $\sigma=3$ detection level we are select too few blue ELG, that do in fact show H$\alpha$ emission (seen in the GMOS spectra). This is a consequence of the different brightness of red and blue galaxies: the mean and standard deviation values of the \rb \ magnitudes is $20.37 \pm 0.82$~mag for red galaxies and $21.58 \pm 0.89$~mag for blue galaxies. This also produces the difference between the S/N ratios at the H$\alpha$ wavelength range for the red and blue galaxies. The median value of the S/N of the three filters closer to H$\alpha$ at the cluster redshift is 14.67 for red galaxies and 5.27 for blue galaxies, which is almost 3 times better for red galaxies than for blue galaxies.

\section{ELG classification}
\label{sec:ELG_class}

\begin{figure*}
    \centering
    \includegraphics[width=0.95\textwidth]{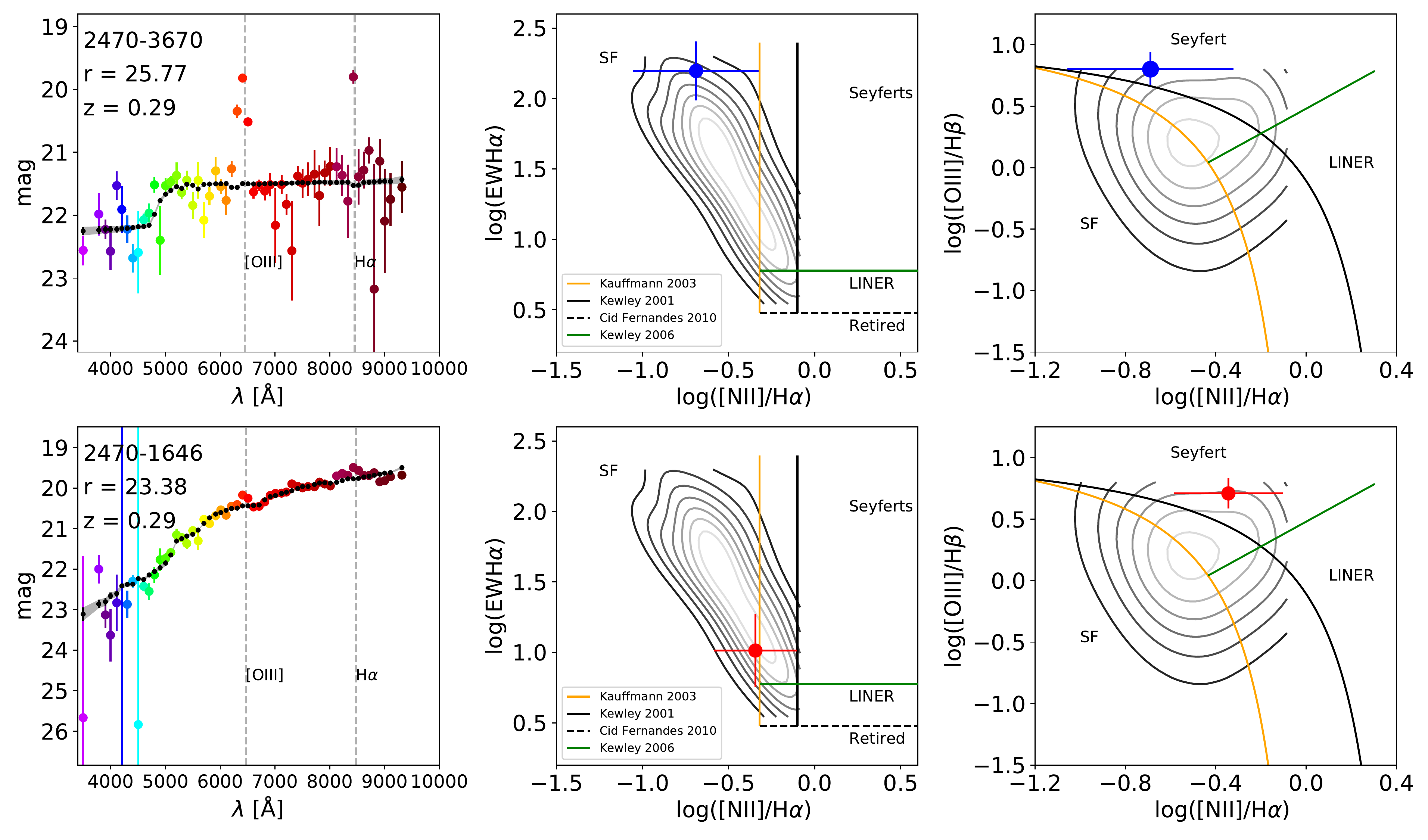}
    \caption{Example of a star-forming galaxy (top row) and a Seyfert galaxy (bottom row). From left to right, the panels in each galaxy show the spectra and its position in the WHAN and BPT diagrams (where the contours represent this work's galaxy density in each diagram). The solid orange and black lines represent the \cite{WHAN_1} transposition of the \cite{Kauffmann2003} and \cite{Kewley2001} SF-AGN distiction criteria, and the green solid line represents the transposition of the \cite{Kewley2006} made by \cite{WHAN_1}. The dashed black line represents the distinction between retired galaxies and LINERs \citep{WHAN_2}.}
    \label{fig:Examples_AGN}
\end{figure*}

\begin{table*}[]
    \centering
    \caption{Classification of the ELG population attending to the WHAN and BPT diagrams. First column shows the ID of the galaxy. Second and third column show the  classification on the WHAN and BPT diagrams respectively. Fourth column shows our final consideration using both diagrams and taking into account the errors of the predictions. Question marks indicate the possibility of the galaxy belonging to such category, but express our insecurity due to errors being to big, the spectrum being too noisy or to a great discrepancy in the WHAN and BPT classes.}
    \begin{tabular}{c c c c }
    \hline
    \hline
        ID & WHAN & BPT & Final classification  \\
        \hline
2470-1030 & SF-Seyfert & Seyfert & AGN\\
2470-1117 & SF & SF-Seyfert & SF\\
2470-1168 & SF & SF & SF\\
2470-1174 & SF & SF-Seyfert & SF\\
2470-1205 & SF-Seyfert & Seyfert & AGN\\
2470-1287 & SF & SF-Seyfert & SF\\
2470-1344 & SF & SF & SF\\
2470-1376 & SF & SF-Seyfert & SF\\
2470-1401 & SF & SF & SF\\
2470-1457 & SF & SF & SF\\
2470-1478 & SF-Seyfert & SF-Seyfert & SF-AGN?\\
2470-1506 & SF & SF & SF\\
2470-1587 & SF-Seyfert & Seyfert & AGN\\
2470-1593 & SF-LINER & Seyfert & AGN\\
2470-1646 & SF & Seyfert & AGN\\
2470-1650 & SF & SF-Seyfert & SF\\
2470-1695 & LINER & SF-LINER & AGN\\
2470-1757 & SF & SF-Seyfert & SF\\
2470-1771 & LINER & LINER & AGN\\
2470-1789 & SF & SF & SF\\
2470-1827 & SF & SF & SF\\
2470-1941 & SF & SF & SF\\
2470-2129 & SF & SF-Seyfert & SF\\
2470-2328 & SF & SF-Seyfert & SF\\
2470-2350 & SF & SF & SF\\
2470-2446 & SF & SF & SF\\
2470-2493 & SF & SF & SF\\
2470-2524 & SF & SF & SF\\
2470-2693 & SF-Seyfert & SF-LINER & SF-AGN?\\
2470-2791 & SF & SF & SF\\
2470-2799 & SF & SF & SF\\
2470-2832 & SF-LINER & Seyfert & AGN\\
2470-2910 & SF & SF-Seyfert & SF\\
2470-2949 & SF & SF & SF\\
2470-2964 & SF-LINER & Seyfert & AGN\\
2470-3255 & SF-Seyfert & SF-LINER & SF-AGN?\\
2470-3345 & SF & SF & SF\\
2470-3670 & SF & Seyfert & SF\\
2470-3712 & SF-Seyfert & Seyfert & AGN\\
2470-3848 & SF-Seyfert & SF-Seyfert & SF-AGN?\\
2470-4414 & SF & SF & SF\\
2470-4691 & SF-Seyfert & Seyfert & AGN\\
2470-492 & SF & SF & SF\\
2470-494 & SF & SF & SF\\
2470-5523 & SF & SF & SF\\
2470-587 & SF & SF & SF\\
2470-701 & SF & SF & SF\\
2470-734 & SF-Seyfert & SF-LINER & SF-AGN?\\
2470-823 & SF-Seyfert & Seyfert & AGN\\
\hline
    \end{tabular}
    \label{tab:AGN}
\end{table*}

In Sect.~\ref{sec:AGN} we described how we classified the ELG into SF galaxies and AGNs using the WHAN and BPT diagrams, and explained that it is difficult to uniquely classify each galaxy. Figure~\ref{fig:Examples_AGN} illustrates this difficulty and the uncertainty  that we face in this classification. 

Fig.~\ref{fig:Examples_AGN} shows two galaxies that are classified as SF in the WHAN diagram and as Seyfert galaxies in the BPT. The \js{} of the galaxy 2470--3670 shows that it is probably an Extreme Emission Line Galaxy (EELG) (Iglesias et al. 2022 submitted), due to its weak continuum and its strong emission lines. However, its Seyfert classification is poorly constrained by [NII]/H$\alpha$ ratio and error. However, its probability to be SF galaxy in the WHAN diagram is equal to one. This is why this galaxy is classified as SF. On the contrary, 2470--1646 is classified well constrained in the BPT as a Seyfert. Also its GMOS spectrum shows [OIII]/H$\beta$ and [NII] /H$\alpha$ ratios of Seyfert galaxies. At the WHAN diagram the \js{} data play it close to the SF region, and the probability to be classified as a SF is 0.55. A deep inspection of the \js{} images at the $H\alpha$ and close continuum filters shows that it is a spiral galaxy with extended H$\alpha$ emission that could be produced by the AGN and young stars. This galaxy is finally classified as an AGN, although it has SF properties. 

These cases motivates our simpler classification according with their probability in the WHAN diagram as SF or BPT diagram as AGN. Other cases that are between the "composite" regions in the WHAN or BPT are classified as SF-AGN.  Table~ \ref{tab:AGN} shows the position on each diagram for every galaxy and the final classification given to it, attending to the criteria described in Sect.~\ref{sec:AGN}.

\end{appendix}

\end{document}